\documentclass[aps,prl,reprint]{revtex4-1}

\usepackage{graphicx}
\usepackage{color}
\usepackage{amsmath}
\usepackage{amsfonts}
\usepackage{amssymb}
\usepackage{dcolumn}
\usepackage{hyperref}
\usepackage{enumerate}
\usepackage{bbold}
\usepackage{cancel}
\usepackage{comment}
\usepackage[capitalize]{cleveref}

\definecolor{g}{rgb}{0.1, 0.5, 0.1}

\begin{document}
  \title{Critical behaviors in contagion dynamics}

 \author{L. B\"{o}ttcher}
  \email{lucasb@ethz.ch}

\author{J. Nagler}
\email{jnagler@ethz.ch}
 \affiliation{ETH Zurich, Wolfgang-Pauli-Strasse 27, CH-8093 Zurich,
Switzerland}
 \author{H. J. Herrmann}
  \affiliation{ETH Zurich, Wolfgang-Pauli-Strasse 27, CH-8093 Zurich,
Switzerland}  
\date{\today}
\begin{abstract} 
We study the critical behavior of 
a general contagion model where nodes are 
either active (e.g.~with opinion A, or functioning) or inactive (e.g.~with opinion B, or damaged).
The transitions between these two states are determined by
(i) spontaneous transitions independent of the neighborhood, 
(ii) transitions induced by neighboring nodes and
 (iii) spontaneous reverse transitions.
  The resulting dynamics 
  is extremely rich
  including
  limit cycles and random phase switching.
We
derive
a unifying mean-field theory.
Specifically, we 
analytically 
show that 
the critical behavior of systems whose dynamics is governed by processes (i-iii) 
can only exhibit three distinct regimes:
(a)
uncorrelated spontaneous transition dynamics
(b) contact process dynamics and
(c) cusp catastrophes.
This ends a long-standing debate on the universality classes of complex contagion dynamics in mean-field
and substantially deepens its mathematical understanding. 
\end{abstract}
\maketitle
In 1972 \emph{Schl\"{o}gl} proposed two models describing autocatalytic chemical reactions \cite{schloegl1972} 
that are commonly known today as \emph{Schl\"{o}gl's first} and \emph{Schl\"{o}gl's second} model 
henceforth referred to as \emph{Schl\"{o}gl I} and \emph{Schl\"{o}gl II}). 
Schl\"{o}gl I, also known as 
\emph{contact process} (Harris 1974), comprises the important case of \emph{simple contagion}, i.e.\;the 
\emph{susceptible-infected-susceptible} (SIS) model 
where 
healthy individuals can be infected due to the exposure to a single infectious source, eventually
leading to  
the spread of an epidemic disease \cite{harris74,marro05,keeling-rohani2008,satorras14}. 
In contrast, 
Schl\"{o}gl II 
that is also known as quadratic contact process \cite{durrett1999}
requires 
contact to two sources.
Later studies on Schl\"{o}gl II sparked a debate on its critical behavior and 
Grassberger
 noticed in 1982 that a relation to the Ising universality class \emph{`would be a most remarkable extension of the universality hypothesis, from models with detailed balance to models without it'} to conclude that Schl\"{o}gl II \emph{`is not an example of universality between models with and without detailed balance'} \cite{grassberger82}. 

Closely related to this debate, but more recently, 
a generalized
model
of Schl\"{o}gl II 
has been proposed where 
an
arbitrary
number of sources is necessary to induce a transition \cite{tome15}.
The study of the model's
mean-field critical behavior
  led the authors to conjecture that such general failure-recovery dynamics 
belong to the Ising universality class \cite{majdandzic14}. 
This model is of particular interest since it not only includes simple contagions but also \emph{complex contagion} phenomena such as the diffusion of innovations \cite{coleman57,rogers2010diffusion}, political mobilization \cite{chwe99} and viral marketing \cite{leskovec07} that require social reinforcement, 
i.e.~the connection to multiple sources \cite{granovetter78, macy07}.
The model displays an intricate and very rich
dynamics
including hysteresis effects, limit cycles
and cusp catastrophes \cite{ludwig78, zeeman79, strogatz14,majdandzic14,valdez16,boettcher162}.
Thus, a unifying mean-field theory
of the critical behavior
 is essential for a broad range of dynamical systems..

However, the relation to contact process dynamics and cusp catastrophes has only been shown for specific values of the model's parameters \cite{boettcher162}.
 But, given the model's parameter regime,
  can we generally predict the dynamics type?
  And 
  does the model's mean-field critical behavior belong to the Ising universality class or not? 
Here we
answer these questions and
analytically demonstrate 
that the mean-field critical behavior of the model
is restricted to only three possible regimes:
 (a) uncorrelated spontaneous transition dynamics,
 (b) contact process dynamics and
  (c) cusp catastrophes. 
  Cusp catastrophes
  can display abrupt transitions and hysteresis effects ---
  phenomena that can harm the proper functioning of
 real-world networked systems since small variations in the system's control parameters may cause catastrophic transitions from a seemingly well-functioning state to global malfunction or severe outages \cite{achlioptas09,araujo10,nagler11,nagler12,schroeder13,cho13,Helbing13,boettcher14,souza15,boettcher16}. 

\paragraph{Model}

The general contagion dynamics is defined in a network whose constituents (i.e.~nodes) are regarded as either active (e.g.~not damaged) or inactive (e.g.~failed). Three fundamental processes define the transitions between these two states \cite{majdandzic14,boettcher162}:
 (i)
 nodes undergo
 a spontaneous transition $A\rightarrow X$ from an active ($A$) to an inactive state ($X$) in a time interval $dt$ with probability $p \text{d}t$,
  (ii) if fewer than or equal to $m$ nearest neighbors of a 
node are active, the
node becomes inactive ($Y$) due to an induced transition, i.e.~$A\rightarrow Y$, with probability $r \text{d}t$ and 
(iii) a spontaneous reverse transition with probability $q \text{d}t$ if $X\rightarrow A$ or probability $q' \text{d}t$ if $Y\rightarrow A$. The inactive states $X$ and $Y$ only differ in their reverse transitions and are equivalent if $q=q'$. Process (ii) describes that a node with degree $k$ can become inactive if its number of inactive neighbors is larger or equal to $k-m$.
Similar to threshold models describing complex contagion phenomena, the threshold $m$ defines the number of contacts to inactive nodes that is necessary to induce a transition as defined by process (ii) \cite{granovetter78,watts02,lopez08,gleeson2013}. 
A low value of $m$ corresponds to the situation where many inactive neighbors are required to sustain spreading.
In contrast, for
a large value of $m$ only a few inactive neighbors can sustain the spreading process.
\begin{figure}
\begin{minipage}{0.4\textwidth}
\centering
\includegraphics[width=\textwidth]{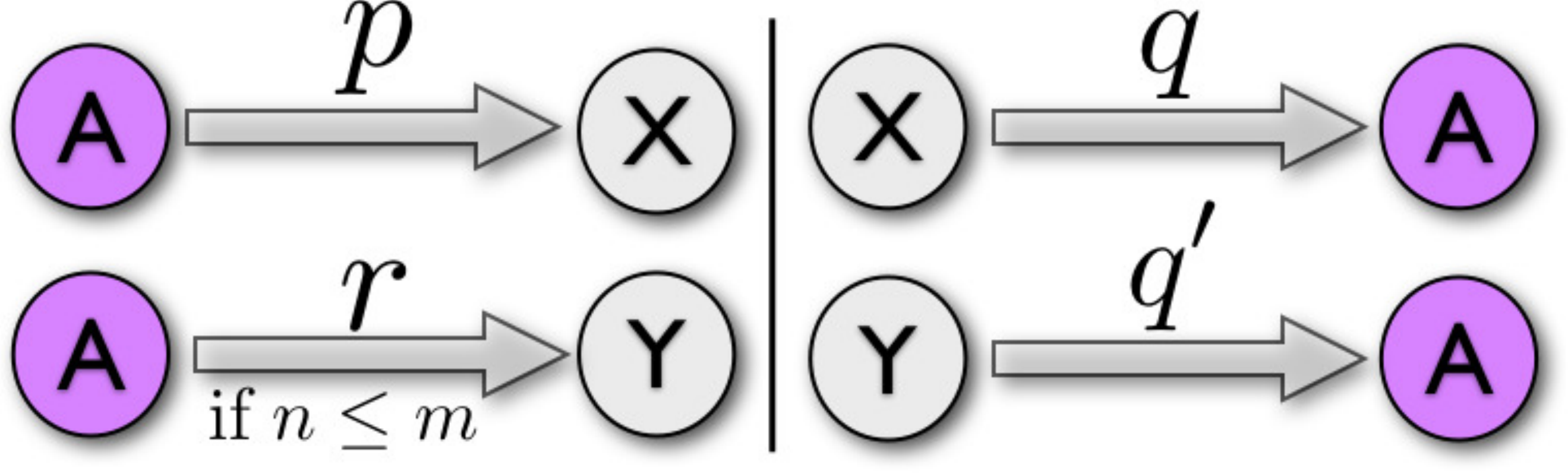}
\end{minipage}
  \caption{\textbf{Model.} Spontaneous failure ($A\rightarrow X$) and spontaneous recovery  ($X\rightarrow A$) occurs with rates $p$ and $q$, respectively. A node may also fail (become inactive) dependent on its neighborhood,
   if too few {\em active} nearest neighbors $n \leq m$ sustain the node's activity ($A\rightarrow Y$ with rate $r$).
  In addition, a failed node $Y$ recovers ($Y\rightarrow A$) with rate $q'$. Active nodes ($A$) are purple while failed ones ($X$ and $Y$) are grey.} 
 \label{fig:model}
\end{figure}
Processes (i-iii) are illustrated in Fig.~\ref{fig:model}.

Let $a(t)\in [0,1]$ denote the total fraction of {\em inactive} nodes. Thus, $a(t)=u_{\text{spon}}(t)+u_{\text{ind}}(t)$ with $u_{\text{spon}}(t)$ and $u_{\text{ind}}(t)$ being the fractions of nodes that are inactive due to spontaneous and induced transitions respectively. The total fraction of inactive nodes in the stationary state is referred to as $a_{st}$. 
In accordance with Ref.~\cite{boettcher162} we derive the mean-field rate equations by assuming a system with homogeneous degrees in the thermodynamic limit that exhibits perfect mixing. Here, perfect mixing either refers to a network of randomly connected nodes with a sufficiently large mean degree or dynamical rewiring  \cite{buckee04,marro05}.
For the fraction of nodes that spontaneously became inactive we find:
\begin{equation}
\dot u_{\text{spon}}
= p \left(1-a
\right)- q \,u_{\text{spon}},
\label{eq:internal_rate}
\end{equation}
where the first term accounts for the fact that active nodes spontaneously become inactive with rate $p$ (process (i)) and the second term corresponds to the spontaneous reverse transition with rate $q$ (process (iii)). Eq.~\eqref{eq:internal_rate} is exact since the network structure is not influencing these spontaneous transitions.

Induced transitions (process (ii)) can only occur for nodes whose number of active neighbors is smaller than or equal to $m$. Under the assumption of a perfectly mixed population, the probability that a node of degree $k$ is located in such a neighborhood is $E_k=\sum_{j=0}^m  \binom {k} {k-j} a^{k-j} (1-a)^j$ \cite{majdandzic14,boettcher162}. The time evolution of the fraction of nodes that are inactive due to induced transitions is therefore given by:
\begin{equation}
\dot u_{\text{ind}}
= r \sum_k f_k E_k \left(1-a\right)- q' u_{\text{ind}},
\label{eq:external_rate}
\end{equation}
with $f_k$ being the degree distribution. The first term describes the occurrence of induced transitions (process (ii)) with rate $r$ of active nodes in a neighborhood where the number of active neighbors is smaller than or equal to $m$ whereas the second term accounts for the spontaneous reverse transition to an active state with rate $q'$ (process (iii)).
In order to study the influence of different threshold values $m$ on the mean-field critical behavior of Eqs.~\eqref{eq:internal_rate} and \eqref{eq:external_rate}, we consider a regular network with degree $k$, 
i.e.~the degree distribution $f_{k'} = \delta_{k k'}$.
 We will demonstrate below that the model defined by processes (i-iii) can only exhibit three different regimes depending on the choice of $m$. It is important to notice that for more general degree distributions the mean-field critical behavior still falls into these classes, see \emph{Supplementary Material}.

The coupled equations \eqref{eq:internal_rate} and \eqref{eq:external_rate} admit oscillatory behavior for $q'>q$ \cite{boettcher162} as a dynamical feature that does not belong to the critical behavior \cite{strogatz14}. The equations describing the critical behavior, i.e.~$\dot u_{spon}= 0$ and $\dot u_{ind}=0$, 
can be 
decoupled by multiplying one of them
with an appropriate constant excluding limit cycles \cite{strogatz14} --- tantamount to
setting $q=q'=1$. This yields
\begin{equation}
\dot a  = f(a,r,p)=r S(a)+p\left(1-a\right)-a,
\label{eq:general_ode}
\end{equation}
with $S(a)=\sum_{j=0}^m  \binom {k} {k-j} a^{k-j} (1-a)^{j+1}=(1-a)E_k$. We use $S(a)$ as shorthand notation for the probability that an active node is located in a neighborhood that is able to induce a transition. Thus, differences in the inactive states $X$ and $Y$, i.e.~different $q$ and $q'$, do not influence the critical behavior of Eq.~\eqref{eq:general_ode} but only rescale $r$ and $p$. In the following we analyze the stationary states of Eq.~\eqref{eq:general_ode} that will be reached in the long-time limit.

\begin{figure*}
\centering
\includegraphics[width=0.8\textwidth]{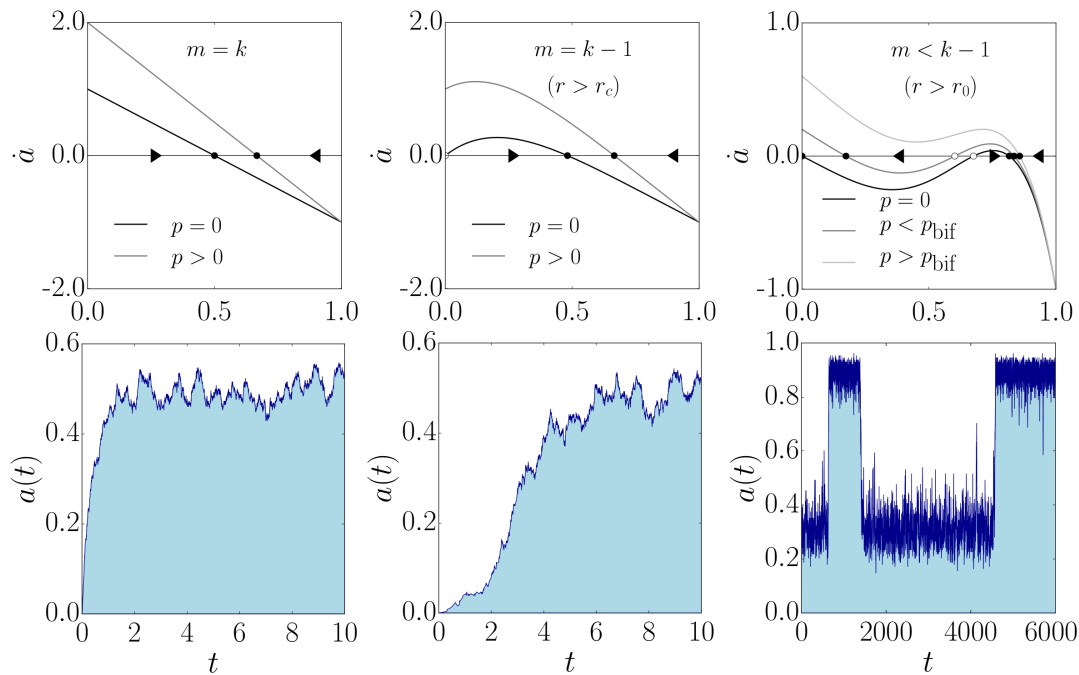}
  \caption{\textbf{Phase portraits of the general contagion model's regimes.} (top panel) The phase portrait is shown for $k=4$ and different values of $m$ with arrows indicating the sign of $\dot{a}$ (right arrow: $\dot{a} > 0$, left arrow: $\dot{a}<0$). Black circles correspond to stable fixed points and white circles to unstable ones. 
  (left) For $m=k$ there 
  exists only one stable fixed point, whose position depends on $p$. We set $r=1$ and for the grey solid line $p=1$.
  (center) For $m=k-1$ the dynamics resembles the phase space of a contact process with a stable non-zero fixed point for $r>r_c$, where $r_c$ defines the critical spreading rate below which $a$ approaches zero if $p=0$. For $p>0$ and $r>r_c$ the second-order phase transition gets smeared out. We set $r=1$ and for the grey solid line $p=1$. 
   (right) If $m<k-1$, $r$ exceeding $r_0$ and $0<p<p_{\text{bif}}$ (inside the hysteresis region enclosed by the bifurcation lines) 
    it is possible to find two stable fixed points. For $p>p_{\text{bif}}$ the dynamics exhibits only one stable fixed point. We set $r=10$ and $m=0$. For the grey solid line ($p<p_{\text{bif}}$) we set $p=0.2$ and for silver solid line ($p>p_{\text{bif}}$) we used $p=0.6$.} (bottom panel) The time evolution for different parameters in a regular random graph with $N=256$ nodes and $k=4$. (left) For $m=k=4$, $p=0.01$, $r=1.0$ the dynamics grows until a stationary state is reached. (center) For $m=k-1=3$, $p=0.01$, $r=1.0$ we find the typical logistic growth pattern. (right) For $m<k-1=1$, $p=0.24$, $r=10$ we encounter phase-switching.
 \label{fig:fixed_points}
\end{figure*}

\paragraph{Class (a): Uncorrelated spontaneous transitions}

We start with the case $m=k$ where the number of active nodes necessary to sustain spreading has to be smaller or equal to the node's degree $k$ according to the definition of process (ii). This describes the regime where 
spreading occurs 
independently of the neighborhood's state such as in
exogenously driven adoption dynamics
\cite{crane2008,fakhteh14},
\begin{equation}
\dot a = (r+p)\left(1-a\right)-a
\label{eq:exogenous}
\end{equation}
since $E_k=1$ and $S(a)=1-a$.
 Eq.~\eqref{eq:exogenous} has only one stationary state, i.e.\;$a_{st}(r,p)=(r+p)/(1+r+p)$, see Fig.~\ref{fig:fixed_points} (left).

\paragraph{Class (b): Contact dynamics}

\begin{figure}
\begin{minipage}{0.49\textwidth}
\centering
\includegraphics[width=0.8\textwidth]{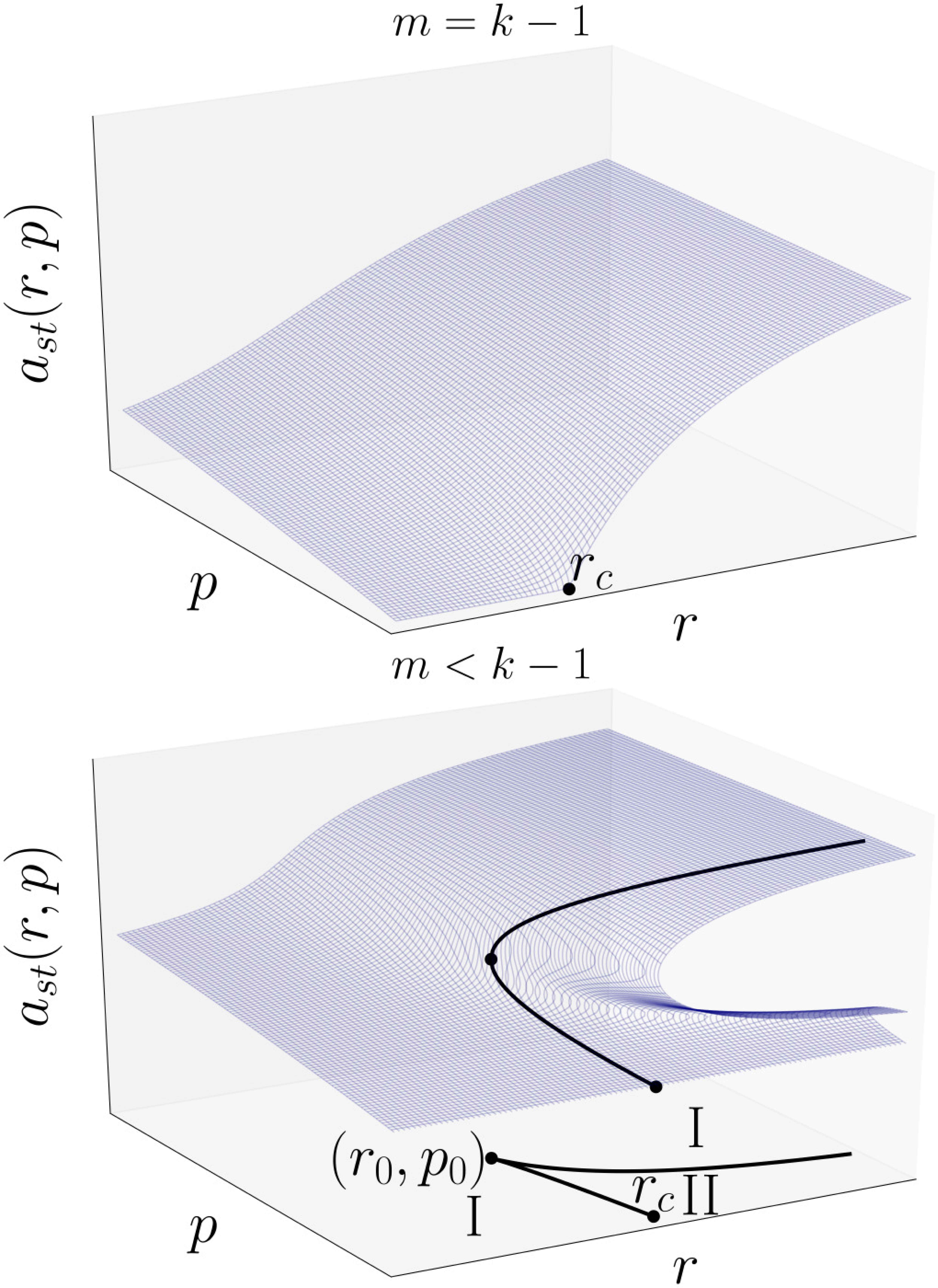}
\end{minipage}
  \caption{\textbf{Phase diagrams of contact dynamics and cusp catastrophes.} (top panel) In the case of $m=k-1$, the phase space corresponds to the one of Schl\"ogl I or contact dynamics. For vanishing $p$, a second-order phase transition from an absorbing to an active phase occurs at $r_c$. For a positive value of $p$, the transition gets smeared out \cite{marro05,henkel08}. (bottom panel) For all $m<k-1$, cusp catastrophes define the phase space. Two in the thermodynamic limit stable states coexist inside the hysteresis region (II) that is surrounded by bifurcation lines (black solid lines) merging at the cusp point $(r_0,p_0)$. Outside the hysteresis region (I) only one stable state exists. The critical point $r_c$ is defined as the transition point for vanishing $p$.}
 \label{fig:phase_spaces}
\end{figure}

By definition $m=k-1$ implies that $k-1$ or less neighbors of a node have to be active to induce a transition. This case describes a contact process where one inactive neighbor is sufficient to sustain spreading \cite{marro05}. As demonstrated in the \emph{Appendix}, we find for $p=0$ that there exists a critical $r_c=k^{-1}$ separating an absorbing and an active phase, i.e.~$a_{st}^{(1)}(r,0)=0$ as $r\leq r_c$ and $a_{st}^{(2)}(r,0)>0$ as $r>r_c$. In the limit of $r\rightarrow r_c$ Eq.~\eqref{eq:general_ode} takes the form:
\begin{equation}
\dot a = r k a  \left(1-a\right)-a.
\label{eq:cp_2}
\end{equation}
Equation (\ref{eq:cp_2}) describes the mean-field contact process, SIS dynamics or Schl\"ogl I \cite{marro05,grassberger82}. In the limit of $r \rightarrow r_c$ the order parameter scales as $a_{st}(r,0) \sim |r-r_c|^\beta$ with $\beta=1$ and adding the field-like contribution $p (1-a)$ to Eq.~\eqref{eq:cp_2} yields $a_{st}(r_c,p) \sim p^{1/\delta_h}$ as $p\rightarrow 0$ with the field exponent $\delta_h=2$. We illustrate in Fig.~\ref{fig:fixed_points} (center) the occurrence of only one stable fixed point for $p>0$. This also results in a smeared-out transition for $p>0$ instead of a second-order phase transition for $p=0$ as shown in Fig.~\ref{fig:phase_spaces} (top panel). For $p=0$ and $r>r_c$ one clearly sees that $a_{st}^{(1)}(r,0)$ is an unstable but $a_{st}^{(2)}(r,0)$  a stable fixed point.

\paragraph{Class (c): Cusp catastrophes}

For some values of $m$, we find a metastable region as illustrated in Fig.~\ref{fig:phase_spaces} (right). Inside this hysteresis region two stable fixed points coexist. Phase-switching is observed when fluctuations in systems of finite size push the dynamics close to the unstable fixed point, cf.~Fig.~\ref{fig:fixed_points} (right). Between the switching events the dynamics remains in one of the two phases for some time. 
The
waiting times thus depend on the fluctuation strength and the distance from one phase to the unstable state in the phase portrait, cf.~Fig.~\ref{fig:fixed_points} (right). For $m < k-1$ where either two or more inactive neighbors are necessary to induce a transition, we now show that the corresponding metastable regions always 
exist due to the relation to cusp catastrophes \cite{zeeman79}. For a detailed analytical treatment we refer to the \emph{Appendix}. The cusp point where the two bifurcation lines intersect (cf.~Fig.~\ref{fig:phase_spaces} (right)) is given by $a_0(k,m)=(k-1-m)/(k+1)$ together with the corresponding control parameters
\begin{equation}
 r_0(k,m)=\frac{1}{S(a_0)+S'(a_0)(1-a_0)},
 \label{eq:r0}
\end{equation}
and
\begin{equation}
 p_0(k,m)=\frac{S'(a_0)a_0-S(a_0)}{S(a_0)+S'(a_0)(1-a_0)}.
  \label{eq:p0}
\end{equation}
We illustrate the influence of different values of $m$ on $(r_0,p_0)$ and on the extent of the hysteresis area in the \emph{Appendix}. Studying Eq.~\eqref{eq:general_ode} in the vicinity of $(a_0,r_0,p_0)$, i.e.~setting $a=a_0+\tilde{a}$, $r=r_0+\tilde{r}$, $p=p_0+\tilde{p}$, yields for the Taylor expansion (omitted tilde) 
\begin{equation}
\dot{a}=r S(a_0)+p(1-a_0) +a (r S'(a_0)-p)+\frac{r_0}{6}S'''(a_0) a^3.
\label{eq:thirdorder}
\end{equation}
We thus find by setting $p$ or $r$ to zero respectively and solving for the fixed point of Eq.~\eqref{eq:thirdorder}:
\begin{equation}
a(r)=\left[ -\frac{6 S(a_0)}{r_0 S'''(a_0)}\right]^{1/3} r^{1/3}+\mathcal{O}\left(r^{2/3}\right),
\label{eq:ar_cusp_m}
\end{equation}
\begin{equation}
a(p)=\left[ -\frac{6 \left(1-a_0 \right)}{r_0 S'''(a_0)}\right]^{1/3} p^{1/3}+\mathcal{O}\left(p^{2/3}\right).
\label{eq:ap_cusp_m}
\end{equation}
In previous work, the critical behavior at the cusp point of a regular random network ($k=10$ and $m=4$) has been conjectured to belong to the Ising universality class although by definition the dynamics corresponds to a general contact process \cite{boettcher162}. 
\paragraph{Final remarks}

We find that the critical behavior of the general contagion model as formulated in Eq.~\eqref{eq:general_ode}
does not belong to the Ising universality class but to
exactly three regimes. 
The first regime, $m=k$, corresponds to purely spontaneous failure and recovery
dynamics.
For $m=k-1$ the model recovers the critical behavior of the contact process.
 A cusp catastrophe is found for all $m<k-1$ with the typical critical behavior at the cusp point 
(Eqs.~\eqref{eq:ar_cusp_m} and (\ref{eq:ap_cusp_m})).
This sheds analytical insight into
a broad range of spreading processes that are determined by
the network's connectivity $k$ and the threshold parameter $m$, cf.~examples in Tab.~\ref{tab:table1}.

We have demonstrated that the phase diagram
corresponds to a cusp catastrophe, when two or more inactive nodes are needed to trigger 
induced node-transitions. This scenario typically implies dramatic and uncontrollable global transitions in the network for many systems involving complex contagion dynamics. One could naively expect that it could be beneficial for failure control to design systems such that a component only fails if many of its neighbors already failed, i.e.~delaying the failure dynamics. Our results suggest, however, that this delaying procedure might facilitate uncontrollable transitions, hence achieving exactly the opposite as initially intended. This result agrees well with previous findings on delaying procedures which have been applied to a SIS model \cite{gross2006,scarpino16}.
For low spatial dimensions or highly structured networks, the assumptions of perfect mixing or independent node-to-node interactions are not guaranteed. Still mean-field approximations qualitatively describe a given dynamics \cite{gleeson12,keeling-rohani2008,marro05}, see given examples in the \emph{Appendix}.

Future work should establish the behavior of
transients as a function of threshold parameter $m$ and the 
topology of the network. 
It has been demonstrated that opinions as well as coinfections may spread faster in clustered networks compared to random ones \cite{centola10,dufresne2015}.
This links 
our 
result to the multiple exposure condition in complex contagion phenomena.

In the study of collective behaviors, such as the  adoption of innovations,
the distinction between exogenous and endogenous factors is of great interest but often solely
 based on a contact process-like adoption model \cite{crane2008,fakhteh14}. 
Our results suggest studying these processes within 
 our
more general framework that incorporates contact process-like adoption as one special case
and can account 
for spreading 
that relies on multiple contacts. 

\begin{table}[htb]
\caption{\label{tab:table1} Examples of models and processes that are related to the classes (a-c).}
\begin{ruledtabular}
\begin{tabular}{p{2.5cm}p{2.5cm}p{2.5cm}}
(a) $m=k$ & (b) $m=k-1$ & (c) $m<k-1$ \\
\hline
Exogenous factors influencing adoption of innovations \cite{fakhteh14}$^\ast$ \newline Social response to exogenous factors \cite{crane2008}$^\dagger$ & Schl\"ogl I \cite{schloegl1972,grassberger82} \newline Contact process \cite{harris74,marro05,henkel08}$^\ast$ \newline SIS model \cite{keeling-rohani2008,satorras14}$^\ast$ \newline Reggeon field theory \cite{grassberger79}$^\ast$ \newline Directed percolation \cite{cardy80}$^\ast$ \newline Bass model \cite{bass69,fakhteh14}$^\ast$ & Schl\"ogl II \cite{schloegl1972,grassberger82}$^\ast$ \newline Quadratic contact process \cite{durrett1999}$^\ast$ \newline General contact process \cite{tome15}$^\ast$ \newline Behavioral adoption \cite{centola10}$^\dagger$ \newline Threshold models of complex contagions \cite{granovetter78,watts02,leskovec07,macy07,lopez08,rogers2010diffusion,gleeson2013}$^{\ast\dagger}$ or coordination games \cite{easley2010}$^\dagger$\\
\multicolumn{3}{l}{\textsuperscript{$\ast$}\footnotesize{exact mean-field correspondence}} \\
\multicolumn{3}{l}{\textsuperscript{$\dagger$}\footnotesize{phenomenological correspondence}}
\end{tabular}
\end{ruledtabular}
\end{table}

We acknowledge financial support from the ETH Risk Center (grant 
RC SP 08-15) and ERC Advanced grant number FP7-319968 FlowCCS.
\newpage
\onecolumngrid
\appendix
\section{Appendix}
\subsection{Critical behavior in a regular network}
This section is dedicated to the analytical derivation of the mean-field critical behavior of Eqs.~\eqref{eq:internal_rate} and \eqref{eq:external_rate} in the main manuscript. Thus, we have to focus on the corresponding stationary states, i.e.~$\dot{ u}_{spon}= 0$ and $\dot{u}_{ind}=0$. Both equations can be decoupled by multiplying one of the two with an appropriate constant and adding them up --- the same can be achieved by setting $q=q'=1$ yielding:
\label{sec:critbhex}
\begin{equation}
\dot{a} = f(a,r,p)=r S(a)+p\left(1-a\right)-a,
\label{eq:general_ode1}
\end{equation}
where $S(a)=\sum_{j=0}^{m}\binom{k}{k-j} a^{k-j} \left(1-a\right)^{j+1}$. 
\paragraph{Class (a): Uncorrelated spontaneous transitions}
In the case of $m=k$, i.e.~spontaneous transitions that occur independent of the neighborhood, the governing rate equation is $\dot{a}=(r+p)(1-a)-a$ since $\sum_{j=0}^{k}\binom{k}{k-j} a^{k-j} \left(1-a\right)^{j}=1$. Therefore, the corresponding stationary state is given by $a_{st}(r,p)=(r+p)/(1+r+p)$.
\paragraph{Class (b): Contact dynamics}
For $m=k-1$ we first set $p=0$ and find
\begin{equation}
\dot{a} = r \left[\sum_{j=1}^{k} \binom{k}{j} (-1)^{j+1} a^{j}\right] \left(1-a\right)-a
\label{eq:cp_1_sm}
\end{equation}
since $\sum_{j=0}^{k-1}\binom{k}{k-j} a^{k-j} \left(1-a\right)^{j}=1-\left(1-a\right)^k$.
We directly observe that there is a fixed point $a_{st}^{(1)}(r,0)=0$ at the origin of the phase portrait, cf.~Fig.~\ref{fig:fixed_points} (center). Furthermore, the function $f(a,r,p)$ of Eq.~\eqref{eq:general_ode1} is given by $f(a,r)=r \left[\left(1-a\right)-\left(1-a\right)^{k+1}\right]-a$ and we find $f'(0,r)=r k - 1$. Consequently, $f'(0,r)>0$ if $r > k^{-1}=r_c$ and there exists one maximum $a_{max}(r,k)=1-\left[ \frac{1}{k+1}\left(r^{-1}+1\right)\right]^{1/k}>0$ if $r>r_c$ ($f''(a_{max})<0$). Thus, $a_{st}^{(1)}(r,0)=0$ is the stable fixed point if $r\leq r_c$. For $r>r_c$, $a_{st}^{(1)}(r,0)=0$ is unstable with $f'(0,r)>0$ and a maximum $a_{max}(r,k)$ exists obeying $0<a_{max}(r,k)<1$ ($f(a_{max})>0$). As a consequence of the latter fact and $f(1,r)=-1$, 
a second second fixed point $a_{st}^{(2)}(r,0)\in (a_{max},1)$ exists, that is stable for $r>r_c$ as illustrated in Fig.~\ref{fig:fixed_points}. In the vicinity of the transition point ($r\rightarrow r_c$), we only consider the dominant linear term in the sum of Eq.~\eqref{eq:cp_1_sm} and obtain:
\begin{equation}
\dot{a} = r k a \left(1-a\right)-a.
\label{eq:cp_2_sm}
\end{equation}
As $r \rightarrow r_c$ the order parameter scales as $a_{st}(r,0) \sim |r-r_c|^\beta$ with $\beta=1$ and assuming a non-zero field like term $p (1-a)$ yields $a_{st}(r_c,p) \sim p^{1/\delta_h}$ as $p\rightarrow 0$ with $\delta_h=2$. 
\paragraph{Class (c): Cusp catastrophes}
If $m < k-1$, the critical behavior is described by a cusp catastrophe. The equilibrium point $a_0$ of our dynamical system $\dot{a}=f(a,r,p)$ is said to correspond to a cusp catastrophe \cite{zeeman79} for the parameters $r_0, p_0$ if it satisfies the following conditions \cite{hoppenstedt12}:
\begin{enumerate}[(i)]
\item The point $a_0$ is a non-hyperbolic equilibrium.
\item The quadratic term of the function $f$ vanishes but not the cubic one, i.e.
\begin{equation*}
\frac{\partial^2 f(a,r,p)}{\partial a^2}\biggr\rvert_{(a,r,p)=(a_0,r_0,p_0)}=0~\text{and}~\frac{\partial^3 f(a,r,p)}{\partial a^3}\biggr\rvert_{(a,r,p)=(a_0,r_0,p_0)}\neq 0.
\end{equation*}
\item The two vectors
\begin{equation*}
\mathbf{v} =
\left(
\begin{array}{c}
\frac{\partial f}{\partial r}\\
\frac{\partial f}{\partial p}\\
\end{array}
\right) 
~\text{and}~
\mathbf{w} =
\left(
\begin{array}{c}
\frac{\partial^2 f}{\partial r \partial a}\\
\frac{\partial^2 f}{\partial p \partial a}\\
\end{array}
\right)
\end{equation*}
are linearly independent.
\end{enumerate}
We start with condition (ii) and find that the vanishing second derivative at $a_0$ implies:
\begin{align}
\begin{split}
& \sum_{j=0}^m \binom{k}{k-j} \left(1-a_0\right)^{j} a_0^{-j} \left[j \left(j+1\right) + \left(a_0-1\right) \left(1+a_0+2j\right)k+\left(a_0-1\right)^2 k^2\right]\\
& =\binom{k}{k-1-m} \left(1-a_0\right)^{m+1} a_0^{-m}\left(1+m\right) \left[k (1-a_0)-1-a_0-m\right]=0.
\end{split}
\end{align}
The cases $a_0=0$ or $a_0=1$ do not correspond to an equilibrium point of Eq.~\ref{eq:general_ode}. We therefore focus on the remaining solution $a_0(k,m)=(k-1-m)/(k+1)$.
The third derivative at this point is given by:
\begin{align*}
\frac{\partial^3 f(a,r,p)}{\partial a^3}|_{(a,r,p)=(a_0,r_0,p_0)} =-\frac{(1 + k)^3(1 + m)  \left(\frac{k-1- m}{k+1}\right)^{k - m}\left(\frac{2 + m}{1 + k}\right)^m \binom{k}{k-1-m}}{(k-1- m)^2}<0,
\end{align*}
and clearly non-vanishing since $m<k-1$. In order to satisfy condition (i), i.e.~$f(a_0,r_0,p_0)=f'(a_0,r_0,p_0)=0$, we find:
\begin{equation}
 r_0(k,m)=\frac{1}{S(a_0)+S'(a_0)(1-a_0)},
 \label{eq:r0}
\end{equation}
\begin{equation}
 p_0(k,m)=\frac{S'(a_0)a_0-S(a_0)}{S(a_0)+S'(a_0)(1-a_0)}.
  \label{eq:p0}
\end{equation}
We need to demonstrate that $r_0$ and $p_0$ are well behaved, i.e.~$0<r_0<\infty$ and $0<p_0<\infty$. Therefore, we need to show (1) $S'(a_0)a_0-S(a_0)>0$ and (2) $S(a_0)+S'(a_0)(1-a_0)>0$. The first condition leads to
\begin{align}
& S'(a_0)a_0=\sum_{j=0}^m \binom{k}{k-j} a_0^{k-j}\left(1-a_0\right)^{j+1}\left[ \left(k-j\right)-\frac{a_0}{1-a_0}\left(j+1\right)\right]\\
& > \sum_{j=0}^m \binom{k}{k-j} a_0^{k-j}\left(1-a_0\right)^{j+1}=S(a_0),
\end{align}
what is fulfilled if $\left(k-j\right)-\frac{a_0}{1-a_0}\left(j+1\right)>1$. Since $a_0=(k-1-m)/(k+1)$, we obtain the condition $(k+1)(m+1-j)>m+2$ 
which is always satisfied for $m\in \{0,\dots,k-2\}$ and $j\in\{0,\dots,m \}$. For condition (2) it is sufficient to show that $S'(a_0)>0$ --- a consequence of the previous statement. The remaining condition (iii) for $\mathbf{v}, \mathbf{w}\in\mathbb{R}^2$ is clearly met since
\begin{equation*}
\mathbf{v} =
\left(
\begin{array}{c}
S(a_0)\\
1-a_0\\
\end{array}
\right)~\text{and}~
\mathbf{w} =
\left(
\begin{array}{c}
S'(a_0)\\
-1\\
\end{array}
\right)
\end{equation*}
are linearly independent. We have thus shown that Eq.~\eqref{eq:general_ode1} has an equilibrium point $a_0$ for the parameters $r_0,~p_0$ that corresponds to a cusp catastrophe if $m<k-1$. We study Eq.~\eqref{eq:general_ode1} in the vicinity of $(a_0,r_0,p_0)$ by setting $a=a_0+\tilde{a}$, $r=r_0+\tilde{r}$, $p=p_0+\tilde{p}$ what yields for its Taylor expansion (omitted tilde sign) \cite{hoppenstedt12}:
\begin{equation}
\dot{a}=f(a_0,r_0,p_0)+\mathbf{v}\cdot \lambda + \mathbf{w}\cdot \lambda a+\frac{1}{3 !}r_0 S'''(a_0) a^3 + \text{h.o.t.},
\end{equation}
where $\lambda=(r,p)$ and h.o.t.~describes the higher-order terms in $a$, $r$ and $p$. Neglecting the terms of higher-order, the latter equation can be also written as:
\begin{equation}
\dot{a}=r S(a_0)+p(1-a_0) +a (r S'(a_0)-p)+\frac{r_0}{6}S'''(a_0) a^3.
\label{eq:thirdorder_sm}
\end{equation}
We find the critical behavior in the vicinity of the cusp point by setting $p$ or $r$ to zero respectively and solving for the fixed point of Eq.~\eqref{eq:thirdorder_sm}:
\begin{equation}
a(r)=\left[ -\frac{6 S(a_0)}{r_0 S'''(a_0)}\right]^{1/3} r^{1/3}+\mathcal{O}\left(r^{2/3}\right),
\label{eq:ar_cusp}
\end{equation}
\begin{equation}
a(p)=\left[ -\frac{6 \left(1-a_0 \right)}{r_0 S'''(a_0)}\right]^{1/3} p^{1/3}+\mathcal{O}\left(p^{2/3}\right).
\label{eq:ap_cusp}
\end{equation}
\subsection{Parameter dependence of the hysteresis area}
\label{sec:hysteresis}
\begin{figure}
\begin{minipage}{0.49\textwidth}
\centering
\includegraphics[width=\textwidth]{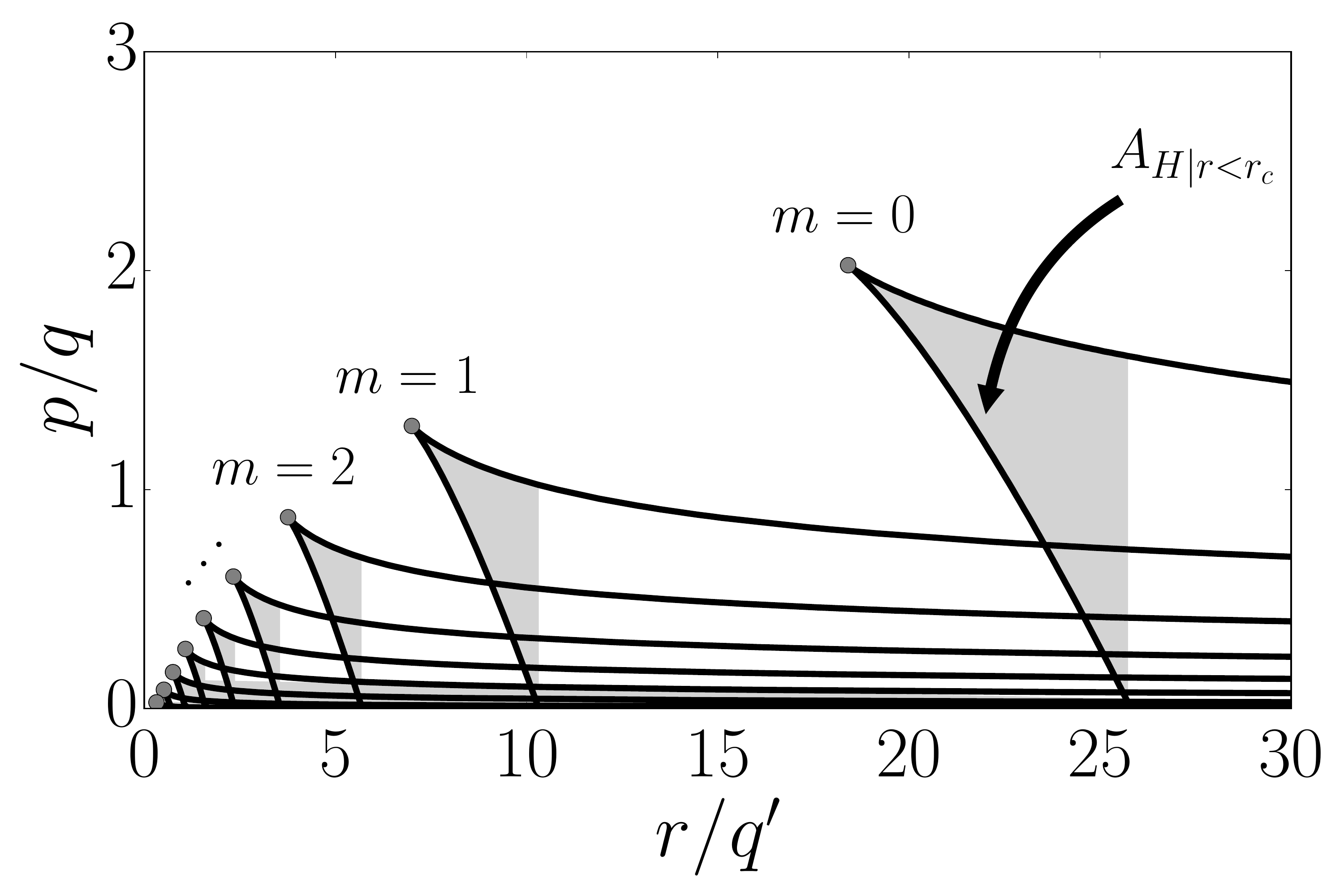}
\end{minipage}
\begin{minipage}{0.49\textwidth}
\centering
\includegraphics[width=\textwidth]{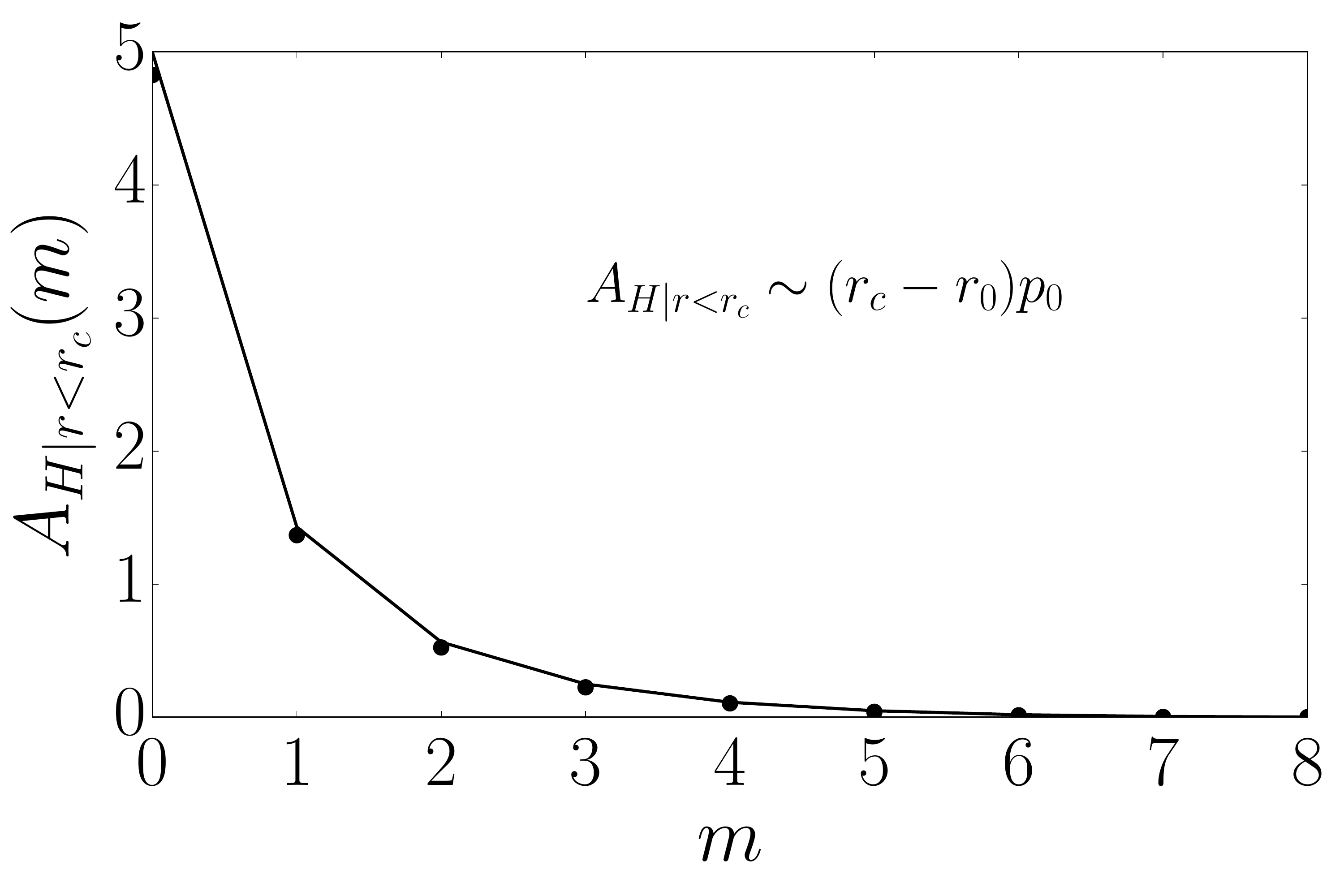}
\end{minipage}
  \caption{\textbf{
  Extent and position of hysteresis regions. 
  } 
  (left) Hysteresis regions for $k=10$ and different $m$. (right) Hysteresis areas for $k=10$ as function of $m$. 
  The black line is the
   approximation
   $A_{H|r<r_c}(r_0,p_0)\approx \frac{1}{3} (r_c-r_0) p_0$ where 
   the $m$-dependence is adopted from 
   the ones of 
   $r_0$ and $p_0$ in Eqs.\;(\ref{eq:r0}) and (\ref{eq:p0}). 
   The parameters $r$ and $p$ are varied whereas $q=q'=1$.} 
 \label{fig:hysteresis_areas}
\end{figure}
Position and extent of the hysteresis region
 impact the
 controllability of failure-recovery dynamics defined by processes (i-iii) \cite{boettcher162}. 
 Here, 
 we study the influence of different values of the threshold $m$ keeping $k$ fixed (Fig.~S\ref{fig:hysteresis_areas}). 
For $k=10$ and $m\in\{0,\dots,8\}$ the grey dots in Fig.~S\ref{fig:hysteresis_areas} (left) are the cusp points described by Eqs.~\eqref{eq:r0} and \eqref{eq:p0} and the following curve describes their position in phase space:
\begin{equation}
\gamma(k,m)=\left(
\begin{array}{c}
r_0(k,m)\\
p_0(k,m)\\
\end{array}
\right).
\end{equation}
The hysteresis area
decreases with $m$.

 This means,
 the more contacts to inactive nodes necessary, the larger the metastable domain and the larger  threshold values,
   the control parameters $r$ and $p$ for which such behavior occurs. 
   We illustrate this effect for hysteresis areas $A_{H|r<r_c}$ restricted to $r<r_c$ in Fig.~S\ref{fig:hysteresis_areas} (right). 
   The black dots represent the actual values of the corresponding area and the black line is an 
   approximation
   based on the assumption that $A_{H|r<r_c}(r_0,p_0)\sim (r_c-r_0) p_0$. 
   The value of $r_c$, as defined in Fig.~\ref{fig:phase_spaces} (right), is determined by Eq.~\eqref{eq:general_ode1} in the limit $p\rightarrow 0$.
\subsection{Critical behavior for general degree distributions}
Any degree distribution leads to a characteristic polynomial of maximum degree three (Eq.~\eqref{eq:thirdorder} in the manuscript). This is, in general, a 
consequence for two-parameter bifurcations of smooth dynamical systems with exactly two control parameters \cite{zeeman79}.
Therefore, our mathematical framework is valid for any choice of $f_k$.
However, as expected, the mean-field theory is not exact
for, e.g., networks with low degree, or networks with a broad degree distribution,
but nevertheless, the maximal degree of the characteristic polynomial necessarily remains invariant under the specific form of the degree distribution and therefore constraints the critical behavior to fall into classes (a-c). Below we describe the treatment of general degree distributions in our mean-field theory and also discuss specific examples --- bearing in mind that the arguments apply for a perfectly mixed mean-field configuration.

The three classes (a-c) we discussed above in Sec.~\emph{Critical behavior in a regular network} are recurring even for more general degree distributions and are useful to qualitatively understand the mean-field dynamics in more general situations. For values of $m$ that depend on the actual node degree, an analogous treatment is possible. In our mean-field theory a general degree distribution different from a regular one ($f_{k'}=\delta_{k k'}$) implies multiple terms in the sum $\sum_{k'} f_{k'}$. A degree distribution with finite cut-off $k_n$ is of the form $f_{k'}=P_{k_1} \delta_{k_1 k'}+P_{k_2} \delta_{k_2 k'}+\dots+P_{k_n} \delta_{k_n k'}$ with normalization $\sum_{k'\in \{k_1,k_2,\dots, k_n\}} P_{k'}=1$. Here $P_{k_i}$ denotes the probability of degree $k_i$. We now define $S_{k_i}(a)=\sum_{j=0}^m  \binom {k_i} {k_i-j} a^{k_i-j} (1-a)^{j+1}=(1-a)E_{k_i}$. Instead of Eq.~\ref{eq:general_ode1}, a general degree distribution as specified above will lead to:
\begin{equation}
\begin{aligned}
\dot{a}=  &P_{k_1} \left[r S_{k_1}(a) + p(1-a) -a \right]+P_{k_2} \left[r S_{k_2}(a) + p(1-a) -a \right] +\\
&P_{k_3} \left[r S_{k_3}(a) + p(1-a) -a \right]+\dots+P_{k_n} \left[r S_{k_n}(a) + p(1-a) -a \right].
\end{aligned}
\label{eq:gen_degree}
\end{equation}
We study the steady states of Eq.~\eqref{eq:gen_degree} to learn about the critical behavior of a general degree distribution. We are now dealing with a sum of terms $P_{k_i}\left[r S_{k_i}(a) + p(1-a) -a \right]$ that are similar to the ones of a regular degree distribution. Therefore, each term corresponds to cusp, contact or spontaneous dynamics, respectively and the more general behavior can be reduced to the ones of a regular network. Depending on the difference $m-k_i$, different phase-space configurations are possible. We describe some possible cases below to illustrate that the three classes of critical behavior of a regular network are recurring and essential to understand this more general configuration.
\begin{figure}
\begin{minipage}{0.49\textwidth}
\centering
\includegraphics[width=\textwidth]{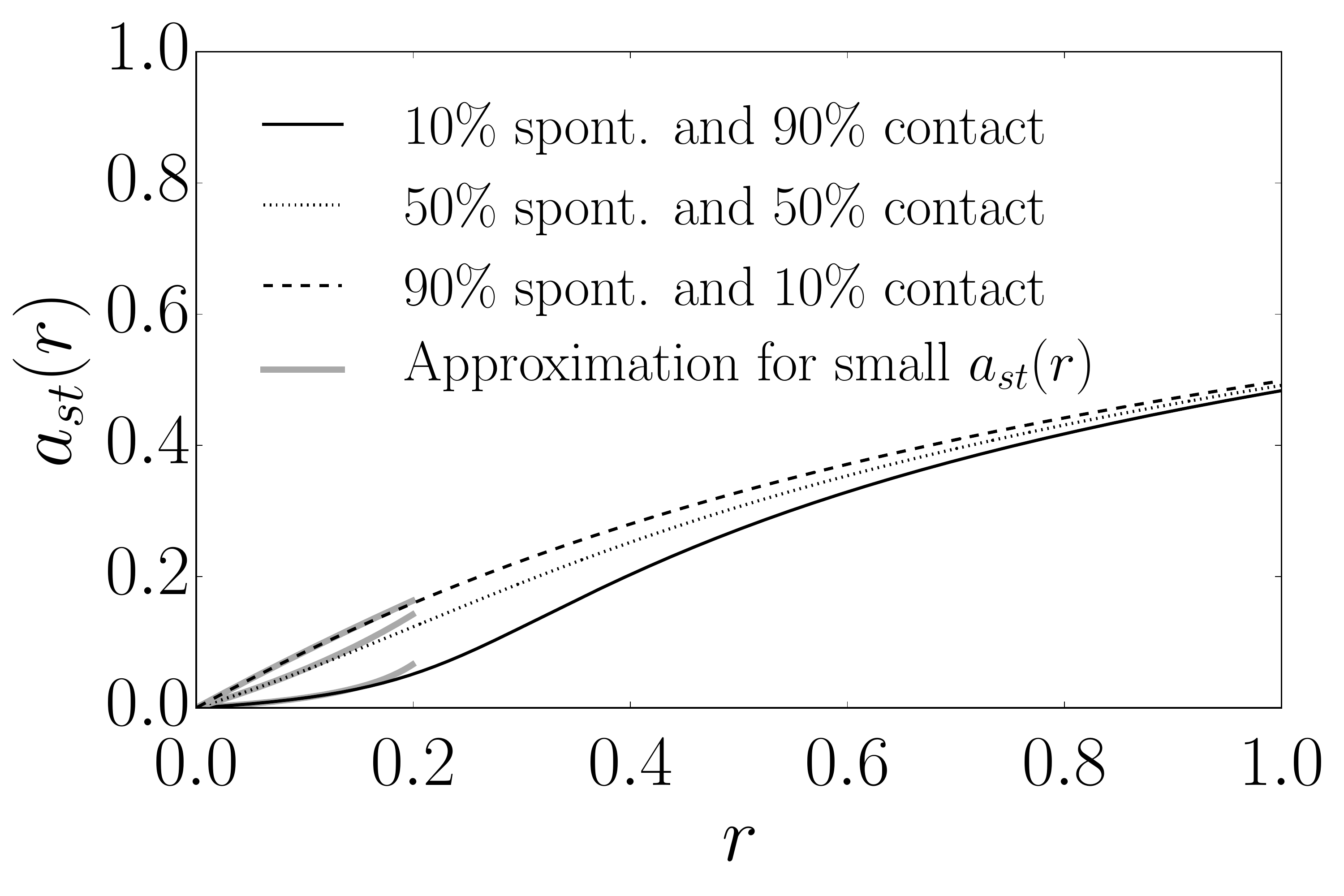}
\end{minipage}
\begin{minipage}{0.49\textwidth}
\centering
\includegraphics[width=\textwidth]{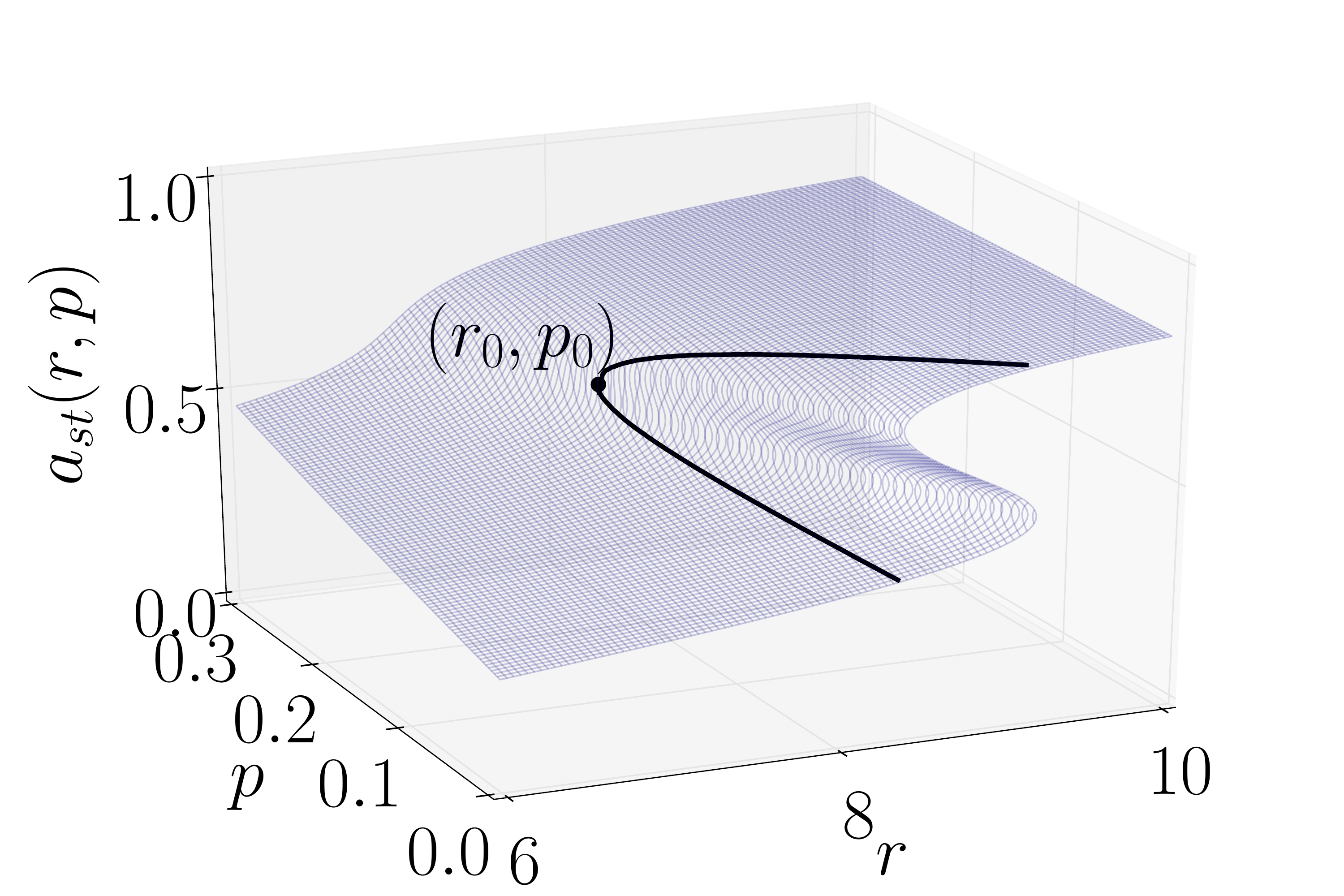}
\end{minipage}
  \caption{\textbf{Examples of mean-field critical behavior for general degree distributions.} (left) An illustration of the growth of the order parameter $a_{st}(r)$ for different mixtures of spontaneous and contact dynamics. The bold grey lines are an approximation for small values of $a_{st}(r)$, cf.~Eq.~\eqref{eq:spont_cont_approx} with $k_i=4$. (right) The phase-space of a mixture of spontaneous and cusp dynamics according to Eq.~\eqref{eq:spont_cusp} with $P_{k_i}=0.95$ and $k_i=4$, $m=0$. The cusp point, as defined by Eqs.~\eqref{eq:r0_spont_cusp} and \eqref{eq:p0_spont_cusp}, is indicated by $(r_0,p_0)$, The black solid lines enclose the hysteresis region.}
 \label{fig:mix_critical}
\end{figure}
\begin{itemize}
\item For $m\geq k_{max}$, the overall dynamics is necessarily spontaneous (as $m\geq k_i$ for all $i$).
\item Spontaneous critical dynamics is observed as well if, for example, the $i$-th term leads to contact dynamics ($m=k_i-1$) and the remaining terms lead to spontaneous ones. Since $m$ is fixed, only one $k_i$ is equal to $m-1$ and with $\sum_{k'\in \{k_1,k_2,\dots, k_n\}\setminus \{k_i\}} P_{k'}=1-P_{k_i}$ (see Sec.~\emph{Critical behavior in a regular network}):
\begin{equation}
0=P_{k_i}\left[r \left(1-(1-a)^{k_i}\right)(1-a)+p(1-a)-a\right]+(1-P_{k_i})\left[(r+p)(1-a)-a\right].
\label{eq:spont_cont}
\end{equation}
A second-order phase transition characterizing the critical behavior of the contact process exists no longer due to spontaneous transitions. For vanishing $p$, the critical behavior of Eq.~\eqref{eq:spont_cont}, i.e.~in this case an expansion for small $a$, yields cf.~Fig.~S\ref{fig:mix_critical}:
\begin{equation}
a_{st}\left(r,p=0,k_i,P_{k_i}\right)=\frac{(1-P_{k_i}) r}{(1-P_{k_i})(1+r)+P_{k_i}(1-k_i r)}+\text{h.o.t.},
\label{eq:spont_cont_approx}
\end{equation}
where $\text{h.o.t.}$ denotes the higher order terms in the occurring variables. We note that Eq.~\eqref{eq:spont_cont_approx} corresponds to the steady states described by Eq.~\eqref{eq:exogenous} for $P_{k_i}= 0$ (degrees $k_i$ are absent), i.e.~$a_{st}(r,p=0)=r/(1+r)$ for spontaneous dynamics in a regular network.
\item Another scenario is the occurrence of a cusp catastrophe for the $i$-th term ($m<k_i-1$) and spontaneous dynamics otherwise. The spontaneous terms just lead to additional constant and linear contributions in the polynomial describing the steady states:
\begin{equation}
0=P_{k_i}\left[r S_{k_i}(a)+p(1-a)-a\right]+(1-P_{k_i})\left[(r+p)(1-a)-a\right].
\label{eq:spont_cusp}
\end{equation}
The cusp point $a_0(k,m)=(k-1-m)/(k+1)$ for regular degree distributions has been derived via a vanishing second order derivative, cf.~Sec.~\emph{Critical behavior in a regular network}. Therefore, the additional constant and linear terms are not affecting the existence of the cusp point in Eq.~\eqref{eq:spont_cusp}. 
As for the regular graph, the non-hyperbolic equilibrium condition yields:
\begin{equation}
 r_0\left(k,m,k_i,P_{k_i}\right)=\frac{1}{P_{k_i}\left[S_{k_i}(a_0)+S_{k_i}'(a_0)(1-a_0)\right]},
 \label{eq:r0_spont_cusp}
\end{equation}
\begin{equation}
 p_0\left(k,m,k_i,P_{k_i}\right)=\frac{P_{k_i}-1+P_{k_i}\left[S_{k_i}'(a_0)a_0-S_{k_i}(a_0)\right]}{P_{k_i}\left[S_{k_i}(a_0)+S_{k_i}'(a_0)(1-a_0)\right]}.
  \label{eq:p0_spont_cusp}
\end{equation}
We note that for $P_{k_i}= 1$ (degrees $k_i$ occur P-almost surely) Eqs.~\eqref{eq:r0_spont_cusp} and \eqref{eq:p0_spont_cusp} correspond to Eqs.~\eqref{eq:r0} and \eqref{eq:p0} for a regular graph. However, unlike Eq.~\eqref{eq:p0}, Eq.~\eqref{eq:p0_spont_cusp} might be a negative, unphysical rate $p_0$. Therefore, the proportion of spontaneous terms determines if the cusp point exists for physical, i.e.~positive rates $r$ and $p$ in the phase-space or not. Consequently, one might not observe a hysteresis region if spontaneous dynamics has too much influence. We illustrate a phase-space configuration for a mixture of spontaneous and cusp dynamics in Fig.~\ref{fig:mix_critical} (right) with $P_{k_i}=0.95$ (degrees $k_i$ occur with probability $0.95$) and $m=0$.
\end{itemize}
\begin{figure}[htp!]
\centering
\begin{minipage}{0.49\textwidth}
\includegraphics[width=\textwidth]{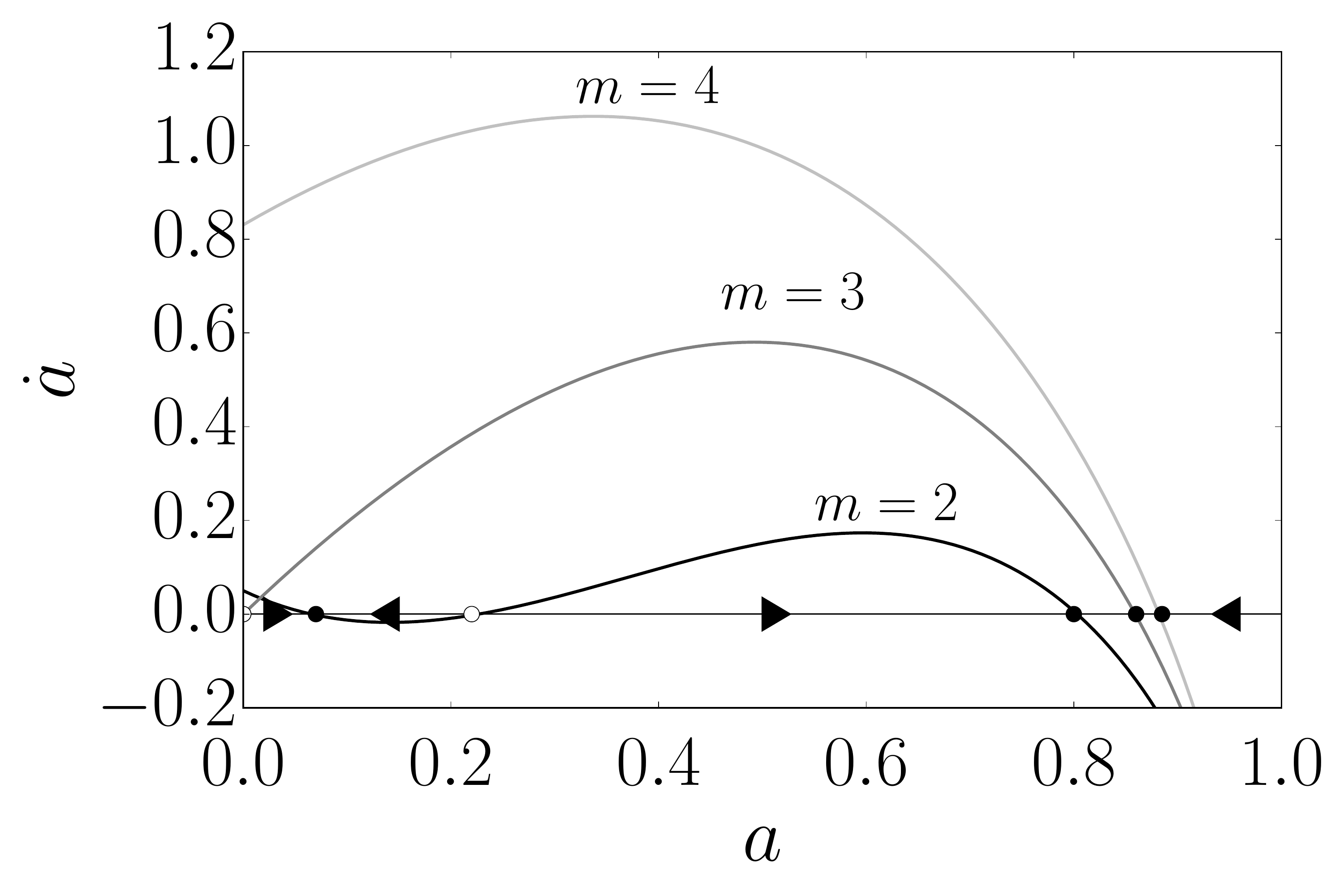}
\end{minipage}
  \caption{\textbf{Phase-portrait configurations of a power-law degree distribution.} The mean-field phase-portrait for a degree distribution $P(k)\propto k^{-1}$ with $k_{min}=4$, $k_{max}=100$, $r=10$ and different values of $m$. For $m=2$ and $m=4$ we set $p=0.05$ and $p=0$ for $m=3$ to illustrate the existence of the unstable fixed point at the origin. Arrows are indicating the sign of $\dot{a}$ (right arrow: $\dot{a} > 0$, left arrow: $\dot{a}<0$). Black circles correspond to stable fixed points and white circles to unstable ones.}
 \label{fig:powerlaw}
\end{figure}
In Fig.~\ref{fig:powerlaw}, we study the influence of a power-law degree distribution $P(k)\propto k^{-1}$ with $k_{min}=4$ and $k_{max}=100$ on the mean-field critical behavior as described by Eq.~\eqref{eq:gen_degree}. We qualitatively recover the three classes describing spontaneous, contact and cusp dynamics. In particular, for $m=2$ we find a polynomial that corresponds to cusp dynamics and for $m=3$ the phase-portrait configuration corresponds to contact dynamics. In the case of $m=4$, the dynamics is spontaneous although the phase-portrait is not linear as in Fig.~(2) (in the manuscript) for a regular network. This is due to the higher order contributions but still there exists only one fixed point and the dynamics is spontaneously driven.
\pagebreak
\subsection{Stochastic simulations}
\begin{figure}[!htp]
\centering
\includegraphics[width=0.5\textwidth]{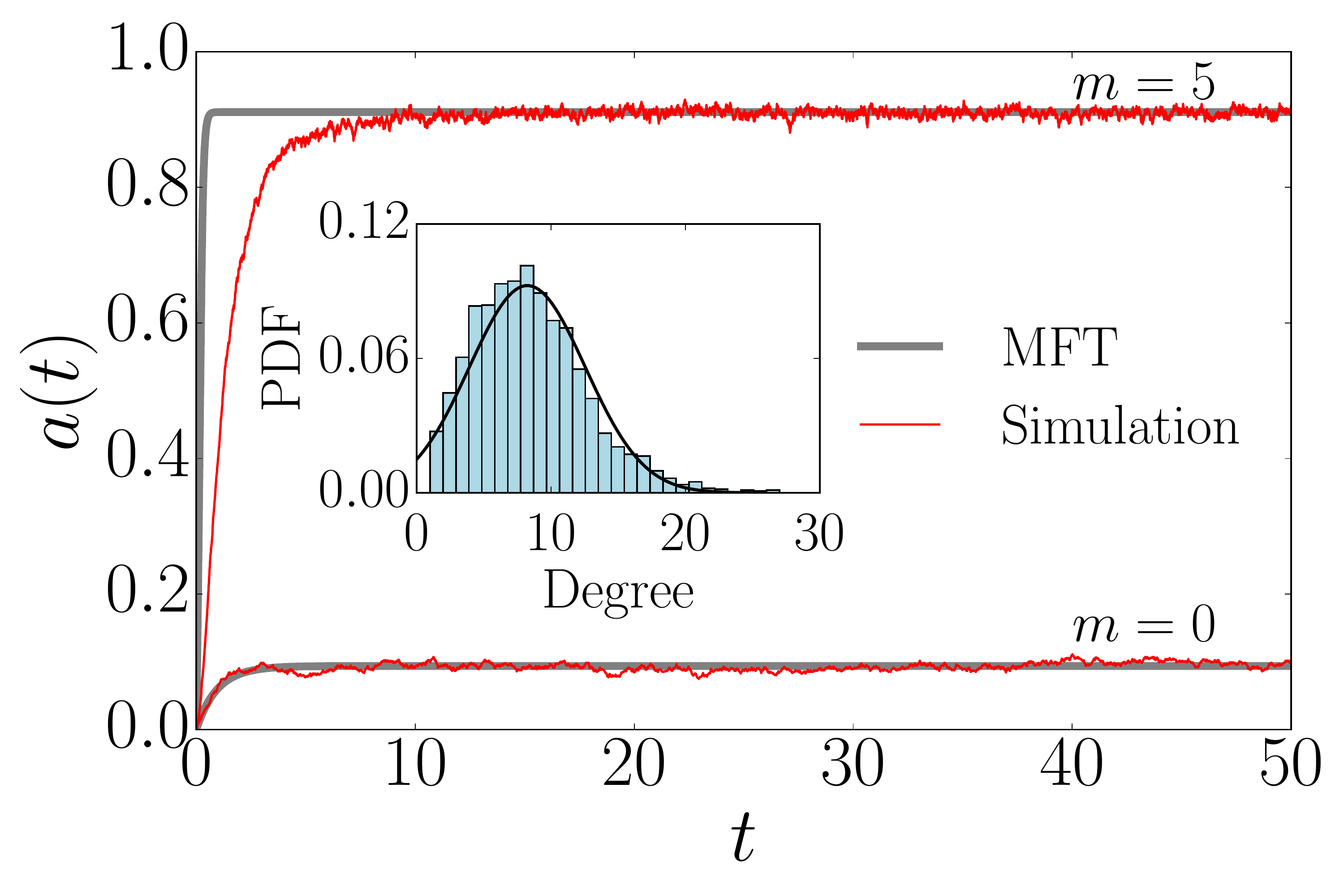}
  \caption{\textbf{Dynamics in a social network.} The time-evolution of the fraction of inactive nodes $a(t)$ for $r=10$, $p=0.1$, $q=q'=1$, $m=0$ and $m=5$ (red solid lines) in a friendship network with 2539 nodes and 20910 edges --- the inset shows its degree distribution. The grey solid lines are the numerical solution of Eqs.~\eqref{eq:internal_rate} and \eqref{eq:external_rate} with the corresponding degree distribution that can be approximated by a truncated normal distribution.}
 \label{fig:school}
\end{figure}

\begin{figure}[!htp]
\begin{minipage}{0.49\textwidth}
\centering
\includegraphics[width=\textwidth]{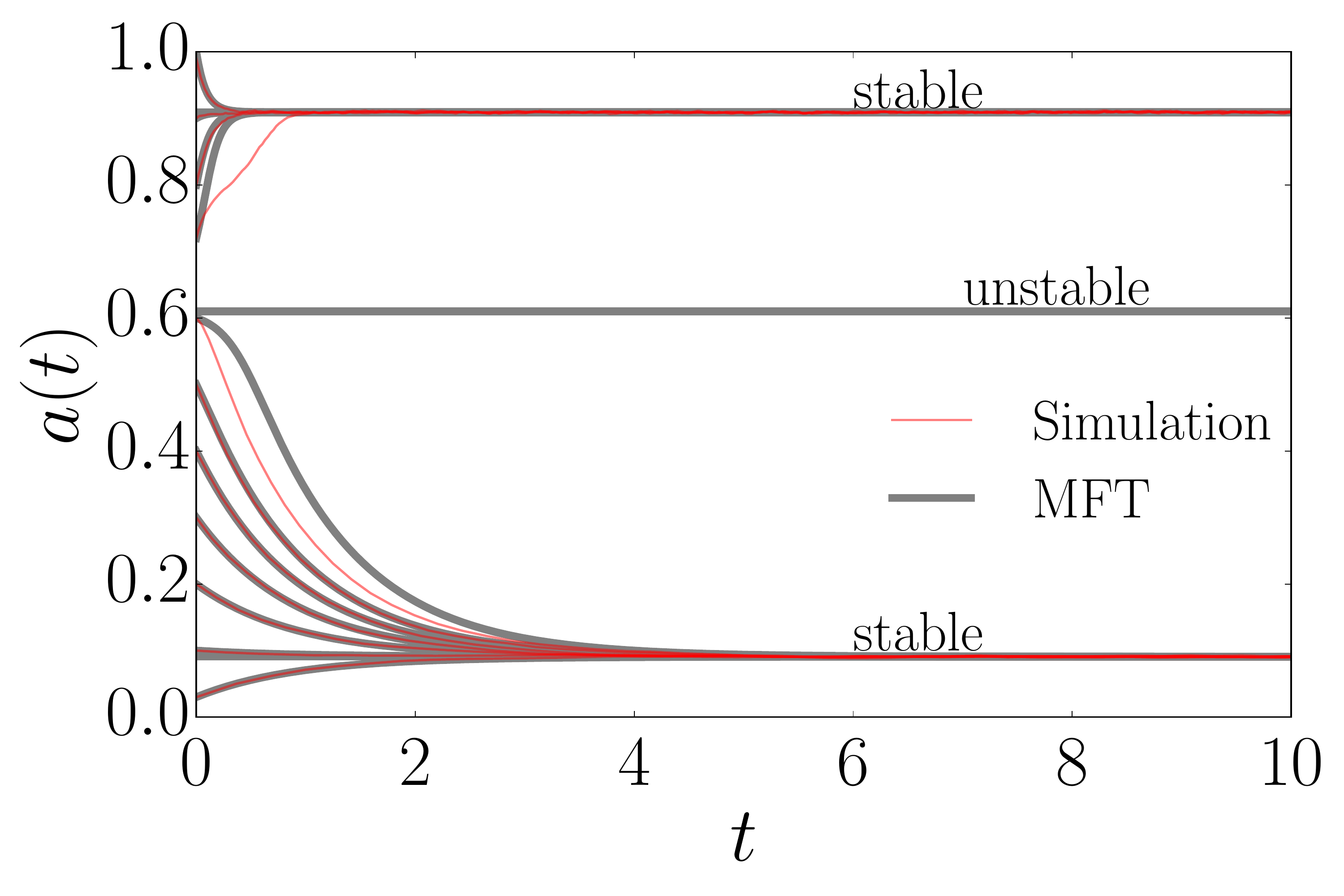}
\end{minipage}
\begin{minipage}{0.49\textwidth}
\centering
\includegraphics[width=\textwidth]{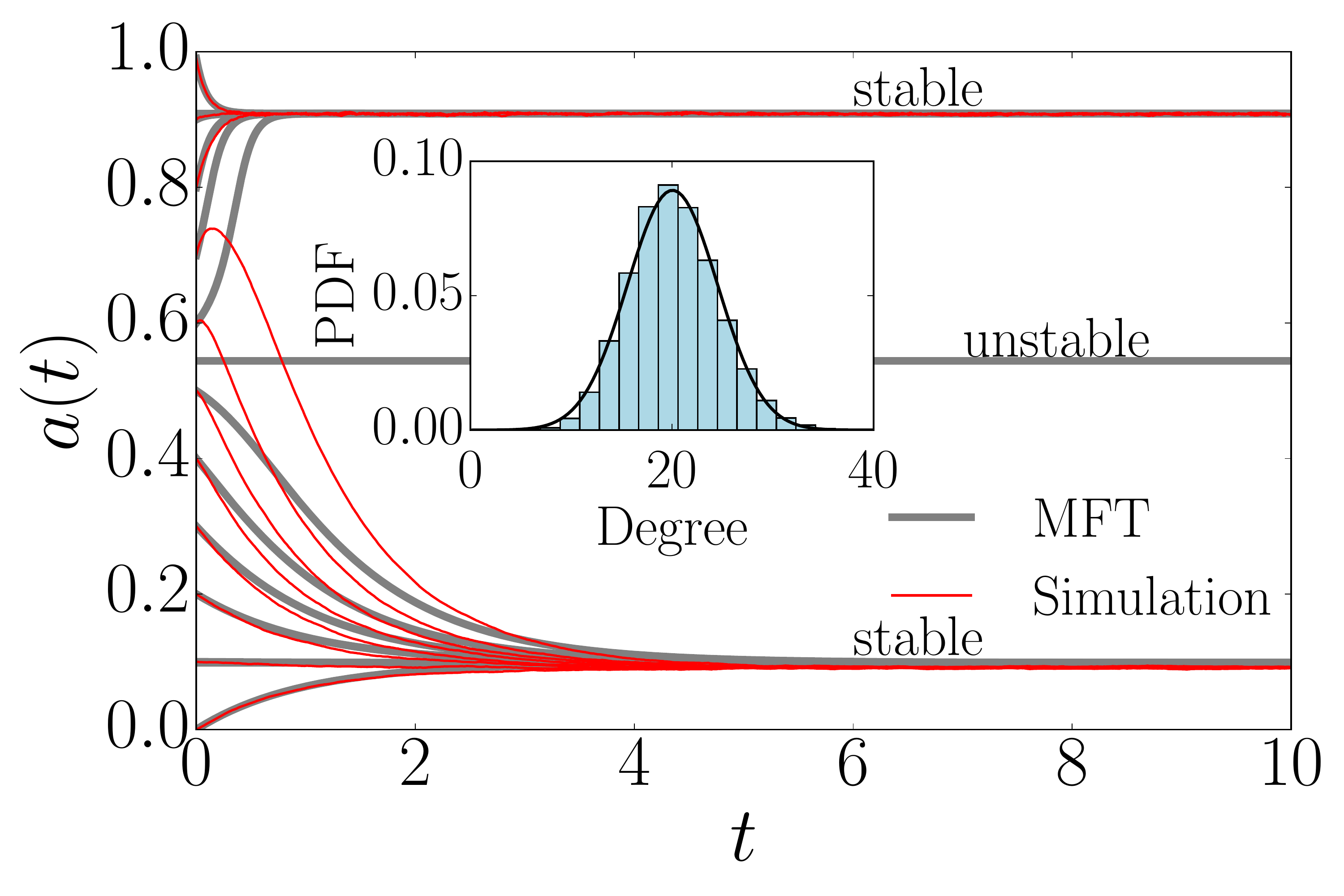}
\end{minipage}
  \caption{\textbf{Stochastic simulation and mean-field theory.} The time-evolution of the fraction of inactive nodes $a(t)$ for different initial conditions $a(t=0)$ in a (left) regular random network with $k=20$  and (right) an Erd\H{o}s-R\'{e}nyi network with average degree $\langle k \rangle = 20$ --- its degree distribution is shown as an inset. Both networks have 100.000 nodes. The red solid lines represent the solutions of the stochastic simulation. The grey solid lines are the corresponding mean-field solutions. Two stable states indicate that the dynamics is located in the hysteresis region. The parameters are $r=10$, $p=0.1$, $q=q'=1$, $m=5$.} 
 \label{fig:stochastic}
\end{figure}

In this section, we present stochastic simulations of spreading processes on real networks compared to the corresponding mean-field predictions. In Fig.~S\ref{fig:school}, we considered a friendship network \footnote{The network is based on the public-use dataset from Add Health, a program project designed by J. Richard Udry, Peter S. Bearman, and Kathleen Mullan Harris, and funded by a grant from the National Institute of Child Health and Human Development (P01-HD31921). For data files from Add Health contact Add Health, Carolina Population Center, 123 W. Franklin Street, Chapel Hill, NC 27516-2524, http://www.cpc.unc.edu/addhealth.} with 2539 nodes and 20910 edges since for example epidemics or opinions spread through such networks. The corresponding degree distribution is illustrated in the inset of Fig.~S\ref{fig:school}. Based on this degree distribution, we numerically computed a mean-field solution (grey solid lines). In Fig.~S\ref{fig:stochastic}, we present simulations of the model's dynamics in a regular random and in an Erd\H{o}s-R\'{e}nyi network.
\newpage
\bibliography{refs}

\begin{thebibliography}{48}%
\makeatletter
\providecommand \@ifxundefined [1]{%
 \@ifx{#1\undefined}
}%
\providecommand \@ifnum [1]{%
 \ifnum #1\expandafter \@firstoftwo
 \else \expandafter \@secondoftwo
 \fi
}%
\providecommand \@ifx [1]{%
 \ifx #1\expandafter \@firstoftwo
 \else \expandafter \@secondoftwo
 \fi
}%
\providecommand \natexlab [1]{#1}%
\providecommand \enquote  [1]{``#1''}%
\providecommand \bibnamefont  [1]{#1}%
\providecommand \bibfnamefont [1]{#1}%
\providecommand \citenamefont [1]{#1}%
\providecommand \href@noop [0]{\@secondoftwo}%
\providecommand \href [0]{\begingroup \@sanitize@url \@href}%
\providecommand \@href[1]{\@@startlink{#1}\@@href}%
\providecommand \@@href[1]{\endgroup#1\@@endlink}%
\providecommand \@sanitize@url [0]{\catcode `\\12\catcode `\$12\catcode
  `\&12\catcode `\#12\catcode `\^12\catcode `\_12\catcode `\%12\relax}%
\providecommand \@@startlink[1]{}%
\providecommand \@@endlink[0]{}%
\providecommand \url  [0]{\begingroup\@sanitize@url \@url }%
\providecommand \@url [1]{\endgroup\@href {#1}{\urlprefix }}%
\providecommand \urlprefix  [0]{URL }%
\providecommand \Eprint [0]{\href }%
\providecommand \doibase [0]{http://dx.doi.org/}%
\providecommand \selectlanguage [0]{\@gobble}%
\providecommand \bibinfo  [0]{\@secondoftwo}%
\providecommand \bibfield  [0]{\@secondoftwo}%
\providecommand \translation [1]{[#1]}%
\providecommand \BibitemOpen [0]{}%
\providecommand \bibitemStop [0]{}%
\providecommand \bibitemNoStop [0]{.\EOS\space}%
\providecommand \EOS [0]{\spacefactor3000\relax}%
\providecommand \BibitemShut  [1]{\csname bibitem#1\endcsname}%
\let\auto@bib@innerbib\@empty
\bibitem [{\citenamefont {Schl\"{o}gl}(1972)}]{schloegl1972}%
  \BibitemOpen
  \bibfield  {author} {\bibinfo {author} {\bibfnamefont {F.}~\bibnamefont
  {Schl\"{o}gl}},\ }\href@noop {} {\bibfield  {journal} {\bibinfo  {journal}
  {Z. Physik}\ }\textbf {\bibinfo {volume} {253}},\ \bibinfo {pages} {147}
  (\bibinfo {year} {1972})}\BibitemShut {NoStop}%
\bibitem [{\citenamefont {Harris}(1974)}]{harris74}%
  \BibitemOpen
  \bibfield  {author} {\bibinfo {author} {\bibfnamefont {T.~E.}\ \bibnamefont
  {Harris}},\ }\href@noop {} {\bibfield  {journal} {\bibinfo  {journal} {Ann.
  Probab.}\ }\textbf {\bibinfo {volume} {2}},\ \bibinfo {pages} {969} (\bibinfo
  {year} {1974})}\BibitemShut {NoStop}%
\bibitem [{\citenamefont {Marro}\ and\ \citenamefont
  {Dickman}(2005)}]{marro05}%
  \BibitemOpen
  \bibfield  {author} {\bibinfo {author} {\bibfnamefont {J.}~\bibnamefont
  {Marro}}\ and\ \bibinfo {author} {\bibfnamefont {R.}~\bibnamefont
  {Dickman}},\ }\href@noop {} {\emph {\bibinfo {title} {{Nonequilibrium Phase
  Transitions in Lattice Models}}}}\ (\bibinfo  {publisher} {Cambridge
  University Press},\ \bibinfo {year} {2005})\BibitemShut {NoStop}%
\bibitem [{\citenamefont {Keeling}\ and\ \citenamefont
  {Rohani}(2008)}]{keeling-rohani2008}%
  \BibitemOpen
  \bibfield  {author} {\bibinfo {author} {\bibfnamefont {M.~J.}\ \bibnamefont
  {Keeling}}\ and\ \bibinfo {author} {\bibfnamefont {P.}~\bibnamefont
  {Rohani}},\ }\href@noop {} {\emph {\bibinfo {title} {{Modeling Infectious
  Diseases in Humans and Animals}}}}\ (\bibinfo  {publisher} {Princeton
  University Press},\ \bibinfo {year} {2008})\BibitemShut {NoStop}%
\bibitem [{\citenamefont {Pastor-Satorras}\ \emph {et~al.}(2015)\citenamefont
  {Pastor-Satorras}, \citenamefont {Castellano}, \citenamefont {Mieghem},\ and\
  \citenamefont {Vespignani}}]{satorras14}%
  \BibitemOpen
  \bibfield  {author} {\bibinfo {author} {\bibfnamefont {R.}~\bibnamefont
  {Pastor-Satorras}}, \bibinfo {author} {\bibfnamefont {C.}~\bibnamefont
  {Castellano}}, \bibinfo {author} {\bibfnamefont {P.~V.}\ \bibnamefont
  {Mieghem}}, \ and\ \bibinfo {author} {\bibfnamefont {A.}~\bibnamefont
  {Vespignani}},\ }\href@noop {} {\bibfield  {journal} {\bibinfo  {journal}
  {Rev. Mod. Phys.}\ }\textbf {\bibinfo {volume} {87}} (\bibinfo {year}
  {2015})}\BibitemShut {NoStop}%
\bibitem [{\citenamefont {Durrett}()}]{durrett1999}%
  \BibitemOpen
  \bibfield  {author} {\bibinfo {author} {\bibfnamefont {R.}~\bibnamefont
  {Durrett}},\ }\href@noop {} {\bibfield  {journal} {\bibinfo  {journal} {SIAM
  Rev.}\ }\textbf {\bibinfo {volume} {41}},\ \bibinfo {pages}
  {677}}\BibitemShut {NoStop}%
\bibitem [{\citenamefont {Grassberger}(1982)}]{grassberger82}%
  \BibitemOpen
  \bibfield  {author} {\bibinfo {author} {\bibfnamefont {P.}~\bibnamefont
  {Grassberger}},\ }\href@noop {} {\bibfield  {journal} {\bibinfo  {journal}
  {Z. Phys. B}\ }\textbf {\bibinfo {volume} {47}},\ \bibinfo {pages} {365}
  (\bibinfo {year} {1982})}\BibitemShut {NoStop}%
\bibitem [{\citenamefont {Tom\'{e}}\ and\ \citenamefont
  {de~Oliveira}(2015)}]{tome15}%
  \BibitemOpen
  \bibfield  {author} {\bibinfo {author} {\bibfnamefont {T.}~\bibnamefont
  {Tom\'{e}}}\ and\ \bibinfo {author} {\bibfnamefont {M.~J.}\ \bibnamefont
  {de~Oliveira}},\ }\href@noop {} {\emph {\bibinfo {title} {{Stochastic
  Dynamics and Irreversibility}}}}\ (\bibinfo  {publisher} {Springer},\
  \bibinfo {year} {2015})\BibitemShut {NoStop}%
\bibitem [{\citenamefont {Majdandzic}\ \emph {et~al.}(2014)\citenamefont
  {Majdandzic}, \citenamefont {Podobnik}, \citenamefont {Buldyrev},
  \citenamefont {Kenett}, \citenamefont {Havlin},\ and\ \citenamefont
  {Stanley}}]{majdandzic14}%
  \BibitemOpen
  \bibfield  {author} {\bibinfo {author} {\bibfnamefont {A.}~\bibnamefont
  {Majdandzic}}, \bibinfo {author} {\bibfnamefont {B.}~\bibnamefont
  {Podobnik}}, \bibinfo {author} {\bibfnamefont {S.~V.}\ \bibnamefont
  {Buldyrev}}, \bibinfo {author} {\bibfnamefont {D.~Y.}\ \bibnamefont
  {Kenett}}, \bibinfo {author} {\bibfnamefont {S.}~\bibnamefont {Havlin}}, \
  and\ \bibinfo {author} {\bibfnamefont {H.~E.}\ \bibnamefont {Stanley}},\
  }\href@noop {} {\bibfield  {journal} {\bibinfo  {journal} {Nat. Phys.}\
  }\textbf {\bibinfo {volume} {10}},\ \bibinfo {pages} {34} (\bibinfo {year}
  {2014})}\BibitemShut {NoStop}%
\bibitem [{\citenamefont {Coleman}\ \emph {et~al.}(1957)\citenamefont
  {Coleman}, \citenamefont {Katz},\ and\ \citenamefont {Menzel}}]{coleman57}%
  \BibitemOpen
  \bibfield  {author} {\bibinfo {author} {\bibfnamefont {J.}~\bibnamefont
  {Coleman}}, \bibinfo {author} {\bibfnamefont {E.}~\bibnamefont {Katz}}, \
  and\ \bibinfo {author} {\bibfnamefont {H.}~\bibnamefont {Menzel}},\
  }\href@noop {} {\bibfield  {journal} {\bibinfo  {journal} {Sociometry}\
  }\textbf {\bibinfo {volume} {20}},\ \bibinfo {pages} {253} (\bibinfo {year}
  {1957})}\BibitemShut {NoStop}%
\bibitem [{\citenamefont {Rogers}(2010)}]{rogers2010diffusion}%
  \BibitemOpen
  \bibfield  {author} {\bibinfo {author} {\bibfnamefont {E.~M.}\ \bibnamefont
  {Rogers}},\ }\href@noop {} {\emph {\bibinfo {title} {{Diffusion of
  Innovations}}}}\ (\bibinfo  {publisher} {Simon and Schuster},\ \bibinfo
  {year} {2010})\BibitemShut {NoStop}%
\bibitem [{\citenamefont {Chwe}(1999)}]{chwe99}%
  \BibitemOpen
  \bibfield  {author} {\bibinfo {author} {\bibfnamefont {M.~S.}\ \bibnamefont
  {Chwe}},\ }\href@noop {} {\bibfield  {journal} {\bibinfo  {journal} {Am. J.
  Sociol.}\ }\textbf {\bibinfo {volume} {105}},\ \bibinfo {pages} {128}
  (\bibinfo {year} {1999})}\BibitemShut {NoStop}%
\bibitem [{\citenamefont {Leskovec}\ \emph {et~al.}(2007)\citenamefont
  {Leskovec}, \citenamefont {Adamic},\ and\ \citenamefont
  {Huberman}}]{leskovec07}%
  \BibitemOpen
  \bibfield  {author} {\bibinfo {author} {\bibfnamefont {J.}~\bibnamefont
  {Leskovec}}, \bibinfo {author} {\bibfnamefont {L.~A.}\ \bibnamefont
  {Adamic}}, \ and\ \bibinfo {author} {\bibfnamefont {B.~A.}\ \bibnamefont
  {Huberman}},\ }\href@noop {} {\bibfield  {journal} {\bibinfo  {journal} {ACM
  Trans. Web}\ }\textbf {\bibinfo {volume} {1}} (\bibinfo {year}
  {2007})}\BibitemShut {NoStop}%
\bibitem [{\citenamefont {Granovetter}(1978)}]{granovetter78}%
  \BibitemOpen
  \bibfield  {author} {\bibinfo {author} {\bibfnamefont {M.}~\bibnamefont
  {Granovetter}},\ }\href@noop {} {\bibfield  {journal} {\bibinfo  {journal}
  {Am. J. Sociol.}\ }\textbf {\bibinfo {volume} {83}},\ \bibinfo {pages} {1420}
  (\bibinfo {year} {1978})}\BibitemShut {NoStop}%
\bibitem [{\citenamefont {Centola}\ and\ \citenamefont {Macy}(2007)}]{macy07}%
  \BibitemOpen
  \bibfield  {author} {\bibinfo {author} {\bibfnamefont {D.}~\bibnamefont
  {Centola}}\ and\ \bibinfo {author} {\bibfnamefont {M.}~\bibnamefont {Macy}},\
  }\href@noop {} {\bibfield  {journal} {\bibinfo  {journal} {Am. J. Sociol.}\
  }\textbf {\bibinfo {volume} {113}},\ \bibinfo {pages} {702} (\bibinfo {year}
  {2007})}\BibitemShut {NoStop}%
\bibitem [{\citenamefont {Ludwig}\ \emph {et~al.}(1978)\citenamefont {Ludwig},
  \citenamefont {Jones},\ and\ \citenamefont {Holling}}]{ludwig78}%
  \BibitemOpen
  \bibfield  {author} {\bibinfo {author} {\bibfnamefont {D.}~\bibnamefont
  {Ludwig}}, \bibinfo {author} {\bibfnamefont {D.~D.}\ \bibnamefont {Jones}}, \
  and\ \bibinfo {author} {\bibfnamefont {C.~S.}\ \bibnamefont {Holling}},\
  }\href@noop {} {\bibfield  {journal} {\bibinfo  {journal} {J. Anim. Ecol.}\
  }\textbf {\bibinfo {volume} {47}},\ \bibinfo {pages} {315} (\bibinfo {year}
  {1978})}\BibitemShut {NoStop}%
\bibitem [{\citenamefont {Zeeman}(1979)}]{zeeman79}%
  \BibitemOpen
  \bibfield  {author} {\bibinfo {author} {\bibfnamefont {E.~C.}\ \bibnamefont
  {Zeeman}},\ }\href@noop {} {\emph {\bibinfo {title} {{Catastrophe theory}}}}\
  (\bibinfo  {publisher} {Springer},\ \bibinfo {year} {1979})\BibitemShut
  {NoStop}%
\bibitem [{\citenamefont {Strogatz}(2014)}]{strogatz14}%
  \BibitemOpen
  \bibfield  {author} {\bibinfo {author} {\bibfnamefont {S.~H.}\ \bibnamefont
  {Strogatz}},\ }\href@noop {} {\emph {\bibinfo {title} {{Nonlinear dynamics
  and chaos: with applications to physics, biology, chemistry, and
  engineering}}}}\ (\bibinfo  {publisher} {Westview press},\ \bibinfo {year}
  {2014})\BibitemShut {NoStop}%
\bibitem [{\citenamefont {Valdez}\ \emph {et~al.}(2016)\citenamefont {Valdez},
  \citenamefont {Muro},\ and\ \citenamefont {Braunstein}}]{valdez16}%
  \BibitemOpen
  \bibfield  {author} {\bibinfo {author} {\bibfnamefont {L.~D.}\ \bibnamefont
  {Valdez}}, \bibinfo {author} {\bibfnamefont {M.~A.~D.}\ \bibnamefont {Muro}},
  \ and\ \bibinfo {author} {\bibfnamefont {L.~A.}\ \bibnamefont {Braunstein}},\
  }\href@noop {} {\bibfield  {journal} {\bibinfo  {journal} {J. Stat. Mech.}\
  }\textbf {\bibinfo {volume} {9}},\ \bibinfo {pages} {093402} (\bibinfo {year}
  {2016})}\BibitemShut {NoStop}%
\bibitem [{\citenamefont {B\"{o}ttcher}\ \emph
  {et~al.}(2016{\natexlab{a}})\citenamefont {B\"{o}ttcher}, \citenamefont
  {Lukovi\'{c}}, \citenamefont {Nagler}, \citenamefont {Havlin},\ and\
  \citenamefont {Herrmann}}]{boettcher162}%
  \BibitemOpen
  \bibfield  {author} {\bibinfo {author} {\bibfnamefont {L.}~\bibnamefont
  {B\"{o}ttcher}}, \bibinfo {author} {\bibfnamefont {M.}~\bibnamefont
  {Lukovi\'{c}}}, \bibinfo {author} {\bibfnamefont {J.}~\bibnamefont {Nagler}},
  \bibinfo {author} {\bibfnamefont {S.}~\bibnamefont {Havlin}}, \ and\ \bibinfo
  {author} {\bibfnamefont {H.~J.}\ \bibnamefont {Herrmann}},\ }\href@noop {}
  {\bibfield  {journal} {\bibinfo  {journal} {arXiv:1610.00997}\ } (\bibinfo
  {year} {2016}{\natexlab{a}})}\BibitemShut {NoStop}%
\bibitem [{\citenamefont {Achlioptas}\ \emph {et~al.}(2009)\citenamefont
  {Achlioptas}, \citenamefont {D'Souza},\ and\ \citenamefont
  {Spencer}}]{achlioptas09}%
  \BibitemOpen
  \bibfield  {author} {\bibinfo {author} {\bibfnamefont {D.}~\bibnamefont
  {Achlioptas}}, \bibinfo {author} {\bibfnamefont {R.~M.}\ \bibnamefont
  {D'Souza}}, \ and\ \bibinfo {author} {\bibfnamefont {J.}~\bibnamefont
  {Spencer}},\ }\href@noop {} {\bibfield  {journal} {\bibinfo  {journal}
  {Science}\ } (\bibinfo {year} {2009})}\BibitemShut {NoStop}%
\bibitem [{\citenamefont {Ara\'{u}jo}\ and\ \citenamefont
  {Herrmann}(2010)}]{araujo10}%
  \BibitemOpen
  \bibfield  {author} {\bibinfo {author} {\bibfnamefont {N.~A.~M.}\
  \bibnamefont {Ara\'{u}jo}}\ and\ \bibinfo {author} {\bibfnamefont {H.~J.}\
  \bibnamefont {Herrmann}},\ }\href@noop {} {\bibfield  {journal} {\bibinfo
  {journal} {Phys. Rev. Lett.}\ }\textbf {\bibinfo {volume} {105}},\ \bibinfo
  {pages} {035701} (\bibinfo {year} {2010})}\BibitemShut {NoStop}%
\bibitem [{\citenamefont {Nagler}\ \emph {et~al.}(2011)\citenamefont {Nagler},
  \citenamefont {Levina},\ and\ \citenamefont {Timme}}]{nagler11}%
  \BibitemOpen
  \bibfield  {author} {\bibinfo {author} {\bibfnamefont {J.}~\bibnamefont
  {Nagler}}, \bibinfo {author} {\bibfnamefont {A.}~\bibnamefont {Levina}}, \
  and\ \bibinfo {author} {\bibfnamefont {M.}~\bibnamefont {Timme}},\
  }\href@noop {} {\bibfield  {journal} {\bibinfo  {journal} {Nat. Phys.}\
  }\textbf {\bibinfo {volume} {7}},\ \bibinfo {pages} {265} (\bibinfo {year}
  {2011})}\BibitemShut {NoStop}%
\bibitem [{\citenamefont {Nagler}\ \emph {et~al.}(2012)\citenamefont {Nagler},
  \citenamefont {Tiessen},\ and\ \citenamefont {Gutch}}]{nagler12}%
  \BibitemOpen
  \bibfield  {author} {\bibinfo {author} {\bibfnamefont {J.}~\bibnamefont
  {Nagler}}, \bibinfo {author} {\bibfnamefont {T.}~\bibnamefont {Tiessen}}, \
  and\ \bibinfo {author} {\bibfnamefont {H.~W.}\ \bibnamefont {Gutch}},\
  }\href@noop {} {\bibfield  {journal} {\bibinfo  {journal} {Phys. Rev. X}\
  }\textbf {\bibinfo {volume} {2}},\ \bibinfo {pages} {031009} (\bibinfo {year}
  {2012})}\BibitemShut {NoStop}%
\bibitem [{\citenamefont {Schr\"{o}der}\ \emph {et~al.}(2013)\citenamefont
  {Schr\"{o}der}, \citenamefont {Rahbari},\ and\ \citenamefont
  {Nagler}}]{schroeder13}%
  \BibitemOpen
  \bibfield  {author} {\bibinfo {author} {\bibfnamefont {M.}~\bibnamefont
  {Schr\"{o}der}}, \bibinfo {author} {\bibfnamefont {S.~H.~E.}\ \bibnamefont
  {Rahbari}}, \ and\ \bibinfo {author} {\bibfnamefont {J.}~\bibnamefont
  {Nagler}},\ }\href@noop {} {\bibfield  {journal} {\bibinfo  {journal} {Nat.
  Commun.}\ }\textbf {\bibinfo {volume} {4}},\ \bibinfo {pages} {2222}
  (\bibinfo {year} {2013})}\BibitemShut {NoStop}%
\bibitem [{\citenamefont {Cho}\ \emph {et~al.}(2013)\citenamefont {Cho},
  \citenamefont {Hwang}, \citenamefont {Herrmann},\ and\ \citenamefont
  {Kahng}}]{cho13}%
  \BibitemOpen
  \bibfield  {author} {\bibinfo {author} {\bibfnamefont {Y.~S.}\ \bibnamefont
  {Cho}}, \bibinfo {author} {\bibfnamefont {S.}~\bibnamefont {Hwang}}, \bibinfo
  {author} {\bibfnamefont {H.~J.}\ \bibnamefont {Herrmann}}, \ and\ \bibinfo
  {author} {\bibfnamefont {B.}~\bibnamefont {Kahng}},\ }\href@noop {}
  {\bibfield  {journal} {\bibinfo  {journal} {Science}\ } (\bibinfo {year}
  {2013})}\BibitemShut {NoStop}%
\bibitem [{\citenamefont {Helbing}(2013)}]{Helbing13}%
  \BibitemOpen
  \bibfield  {author} {\bibinfo {author} {\bibfnamefont {D.}~\bibnamefont
  {Helbing}},\ }\href@noop {} {\bibfield  {journal} {\bibinfo  {journal}
  {Nature}\ }\textbf {\bibinfo {volume} {497}},\ \bibinfo {pages} {51}
  (\bibinfo {year} {2013})}\BibitemShut {NoStop}%
\bibitem [{\citenamefont {B\"{o}ttcher}\ \emph {et~al.}(2015)\citenamefont
  {B\"{o}ttcher}, \citenamefont {Woolley-Meza}, \citenamefont {Ara\'{u}jo},
  \citenamefont {Herrmann},\ and\ \citenamefont {Helbing}}]{boettcher14}%
  \BibitemOpen
  \bibfield  {author} {\bibinfo {author} {\bibfnamefont {L.}~\bibnamefont
  {B\"{o}ttcher}}, \bibinfo {author} {\bibfnamefont {O.}~\bibnamefont
  {Woolley-Meza}}, \bibinfo {author} {\bibfnamefont {N.~A.~M.}\ \bibnamefont
  {Ara\'{u}jo}}, \bibinfo {author} {\bibfnamefont {H.~J.}\ \bibnamefont
  {Herrmann}}, \ and\ \bibinfo {author} {\bibfnamefont {D.}~\bibnamefont
  {Helbing}},\ }\href@noop {} {\bibfield  {journal} {\bibinfo  {journal} {Sci.
  Rep.}\ }\textbf {\bibinfo {volume} {5}},\ \bibinfo {pages} {16571} (\bibinfo
  {year} {2015})}\BibitemShut {NoStop}%
\bibitem [{\citenamefont {D{\rq}Souza}\ and\ \citenamefont
  {Nagler}(2015)}]{souza15}%
  \BibitemOpen
  \bibfield  {author} {\bibinfo {author} {\bibfnamefont {R.~M.}\ \bibnamefont
  {D{\rq}Souza}}\ and\ \bibinfo {author} {\bibfnamefont {J.}~\bibnamefont
  {Nagler}},\ }\href@noop {} {\bibfield  {journal} {\bibinfo  {journal} {Nat.
  Phys.}\ }\textbf {\bibinfo {volume} {11}},\ \bibinfo {pages} {531} (\bibinfo
  {year} {2015})}\BibitemShut {NoStop}%
\bibitem [{\citenamefont {B\"{o}ttcher}\ \emph
  {et~al.}(2016{\natexlab{b}})\citenamefont {B\"{o}ttcher}, \citenamefont
  {Woolley-Meza}, \citenamefont {Goles}, \citenamefont {Helbing},\ and\
  \citenamefont {Herrmann}}]{boettcher16}%
  \BibitemOpen
  \bibfield  {author} {\bibinfo {author} {\bibfnamefont {L.}~\bibnamefont
  {B\"{o}ttcher}}, \bibinfo {author} {\bibfnamefont {O.}~\bibnamefont
  {Woolley-Meza}}, \bibinfo {author} {\bibfnamefont {E.}~\bibnamefont {Goles}},
  \bibinfo {author} {\bibfnamefont {D.}~\bibnamefont {Helbing}}, \ and\
  \bibinfo {author} {\bibfnamefont {H.~J.}\ \bibnamefont {Herrmann}},\
  }\href@noop {} {\bibfield  {journal} {\bibinfo  {journal} {Phys. Rev. E}\
  }\textbf {\bibinfo {volume} {93}},\ \bibinfo {pages} {042315} (\bibinfo
  {year} {2016}{\natexlab{b}})}\BibitemShut {NoStop}%
\bibitem [{\citenamefont {Watts}(2002)}]{watts02}%
  \BibitemOpen
  \bibfield  {author} {\bibinfo {author} {\bibfnamefont {D.~J.}\ \bibnamefont
  {Watts}},\ }\href@noop {} {\bibfield  {journal} {\bibinfo  {journal} {Proc.
  Natl. Acad. Sci.}\ }\textbf {\bibinfo {volume} {99}},\ \bibinfo {pages}
  {5766} (\bibinfo {year} {2002})}\BibitemShut {NoStop}%
\bibitem [{\citenamefont {L\'{o}pez-Pintado}(2008)}]{lopez08}%
  \BibitemOpen
  \bibfield  {author} {\bibinfo {author} {\bibfnamefont {D.}~\bibnamefont
  {L\'{o}pez-Pintado}},\ }\href@noop {} {\bibfield  {journal} {\bibinfo
  {journal} {Game Econ. Behav.}\ }\textbf {\bibinfo {volume} {62}},\ \bibinfo
  {pages} {573} (\bibinfo {year} {2008})}\BibitemShut {NoStop}%
\bibitem [{\citenamefont {Gleeson}(2013)}]{gleeson2013}%
  \BibitemOpen
  \bibfield  {author} {\bibinfo {author} {\bibfnamefont {J.~P.}\ \bibnamefont
  {Gleeson}},\ }\href@noop {} {\bibfield  {journal} {\bibinfo  {journal} {Phys.
  Rev. X}\ }\textbf {\bibinfo {volume} {3}},\ \bibinfo {pages} {021004}
  (\bibinfo {year} {2013})}\BibitemShut {NoStop}%
\bibitem [{\citenamefont {Buckee}\ \emph {et~al.}(2004)\citenamefont {Buckee},
  \citenamefont {Koelle}, \citenamefont {Mustard},\ and\ \citenamefont
  {Gupta}}]{buckee04}%
  \BibitemOpen
  \bibfield  {author} {\bibinfo {author} {\bibfnamefont {C.~O.~F.}\
  \bibnamefont {Buckee}}, \bibinfo {author} {\bibfnamefont {K.}~\bibnamefont
  {Koelle}}, \bibinfo {author} {\bibfnamefont {M.~J.}\ \bibnamefont {Mustard}},
  \ and\ \bibinfo {author} {\bibfnamefont {S.}~\bibnamefont {Gupta}},\
  }\href@noop {} {\bibfield  {journal} {\bibinfo  {journal} {Proc. Natl. Acad.
  Sci.}\ }\textbf {\bibinfo {volume} {101}},\ \bibinfo {pages} {10839}
  (\bibinfo {year} {2004})}\BibitemShut {NoStop}%
\bibitem [{\citenamefont {Crane}\ and\ \citenamefont
  {Sornette}(2008)}]{crane2008}%
  \BibitemOpen
  \bibfield  {author} {\bibinfo {author} {\bibfnamefont {R.}~\bibnamefont
  {Crane}}\ and\ \bibinfo {author} {\bibfnamefont {D.}~\bibnamefont
  {Sornette}},\ }\href@noop {} {\bibfield  {journal} {\bibinfo  {journal}
  {Proc. Natl. Acad. Sci.}\ }\textbf {\bibinfo {volume} {105}},\ \bibinfo
  {pages} {15649} (\bibinfo {year} {2008})}\BibitemShut {NoStop}%
\bibitem [{\citenamefont {Ghanbarnejad}\ \emph {et~al.}(2014)\citenamefont
  {Ghanbarnejad}, \citenamefont {Gerlach}, \citenamefont {Miotto},\ and\
  \citenamefont {Altmann}}]{fakhteh14}%
  \BibitemOpen
  \bibfield  {author} {\bibinfo {author} {\bibfnamefont {F.}~\bibnamefont
  {Ghanbarnejad}}, \bibinfo {author} {\bibfnamefont {M.}~\bibnamefont
  {Gerlach}}, \bibinfo {author} {\bibfnamefont {J.~M.}\ \bibnamefont {Miotto}},
  \ and\ \bibinfo {author} {\bibfnamefont {E.~G.}\ \bibnamefont {Altmann}},\
  }\href@noop {} {\bibfield  {journal} {\bibinfo  {journal} {J. R. Soc.
  Interface}\ }\textbf {\bibinfo {volume} {11}},\ \bibinfo {pages} {20141044}
  (\bibinfo {year} {2014})}\BibitemShut {NoStop}%
\bibitem [{\citenamefont {Henkel}\ \emph {et~al.}(2008)\citenamefont {Henkel},
  \citenamefont {Hinrichsen},\ and\ \citenamefont {L\"{u}beck}}]{henkel08}%
  \BibitemOpen
  \bibfield  {author} {\bibinfo {author} {\bibfnamefont {M.}~\bibnamefont
  {Henkel}}, \bibinfo {author} {\bibfnamefont {H.}~\bibnamefont {Hinrichsen}},
  \ and\ \bibinfo {author} {\bibfnamefont {S.}~\bibnamefont {L\"{u}beck}},\
  }\href@noop {} {\emph {\bibinfo {title} {{Non-Equilibrium Phase Transitions
  Volume I: Absorbing Phase Transitions}}}}\ (\bibinfo  {publisher}
  {Springer},\ \bibinfo {year} {2008})\BibitemShut {NoStop}%
\bibitem [{\citenamefont {Gross}\ \emph {et~al.}(2006)\citenamefont {Gross},
  \citenamefont {D{\rq}Lima},\ and\ \citenamefont {Blasius}}]{gross2006}%
  \BibitemOpen
  \bibfield  {author} {\bibinfo {author} {\bibfnamefont {T.}~\bibnamefont
  {Gross}}, \bibinfo {author} {\bibfnamefont {C.~J.~D.}\ \bibnamefont
  {D{\rq}Lima}}, \ and\ \bibinfo {author} {\bibfnamefont {B.}~\bibnamefont
  {Blasius}},\ }\href@noop {} {\bibfield  {journal} {\bibinfo  {journal} {Phys.
  Rev. Lett.}\ }\textbf {\bibinfo {volume} {96}},\ \bibinfo {pages} {208701}
  (\bibinfo {year} {2006})}\BibitemShut {NoStop}%
\bibitem [{\citenamefont {Scarpino}\ \emph {et~al.}(2016)\citenamefont
  {Scarpino}, \citenamefont {Allard},\ and\ \citenamefont
  {H\'{e}bert-Dufresne}}]{scarpino16}%
  \BibitemOpen
  \bibfield  {author} {\bibinfo {author} {\bibfnamefont {S.~V.}\ \bibnamefont
  {Scarpino}}, \bibinfo {author} {\bibfnamefont {A.}~\bibnamefont {Allard}}, \
  and\ \bibinfo {author} {\bibfnamefont {L.}~\bibnamefont
  {H\'{e}bert-Dufresne}},\ }\href@noop {} {\bibfield  {journal} {\bibinfo
  {journal} {Nat. Phys.}\ }\textbf {\bibinfo {volume} {12}},\ \bibinfo {pages}
  {1042} (\bibinfo {year} {2016})}\BibitemShut {NoStop}%
\bibitem [{\citenamefont {Gleeson}\ \emph {et~al.}(2012)\citenamefont
  {Gleeson}, \citenamefont {Melnik}, \citenamefont {Ward}, \citenamefont
  {Porter},\ and\ \citenamefont {Mucha}}]{gleeson12}%
  \BibitemOpen
  \bibfield  {author} {\bibinfo {author} {\bibfnamefont {J.~P.}\ \bibnamefont
  {Gleeson}}, \bibinfo {author} {\bibfnamefont {S.}~\bibnamefont {Melnik}},
  \bibinfo {author} {\bibfnamefont {J.~A.}\ \bibnamefont {Ward}}, \bibinfo
  {author} {\bibfnamefont {M.~A.}\ \bibnamefont {Porter}}, \ and\ \bibinfo
  {author} {\bibfnamefont {P.~J.}\ \bibnamefont {Mucha}},\ }\href@noop {}
  {\bibfield  {journal} {\bibinfo  {journal} {Phys. Rev. E}\ }\textbf {\bibinfo
  {volume} {85}},\ \bibinfo {pages} {026106} (\bibinfo {year}
  {2012})}\BibitemShut {NoStop}%
\bibitem [{\citenamefont {Centola}(2010)}]{centola10}%
  \BibitemOpen
  \bibfield  {author} {\bibinfo {author} {\bibfnamefont {D.}~\bibnamefont
  {Centola}},\ }\href@noop {} {\bibfield  {journal} {\bibinfo  {journal}
  {Science}\ }\textbf {\bibinfo {volume} {329}},\ \bibinfo {pages} {1194}
  (\bibinfo {year} {2010})}\BibitemShut {NoStop}%
\bibitem [{\citenamefont {H\'{e}bert-Dufresne}\ and\ \citenamefont
  {Althouse}(2015)}]{dufresne2015}%
  \BibitemOpen
  \bibfield  {author} {\bibinfo {author} {\bibfnamefont {L.}~\bibnamefont
  {H\'{e}bert-Dufresne}}\ and\ \bibinfo {author} {\bibfnamefont {B.~M.}\
  \bibnamefont {Althouse}},\ }\href@noop {} {\bibfield  {journal} {\bibinfo
  {journal} {Proc. Natl. Acad. Sci.}\ }\textbf {\bibinfo {volume} {112}},\
  \bibinfo {pages} {10551} (\bibinfo {year} {2015})}\BibitemShut {NoStop}%
\bibitem [{\citenamefont {Grassberger}\ and\ \citenamefont {de~la
  Torre}(1979)}]{grassberger79}%
  \BibitemOpen
  \bibfield  {author} {\bibinfo {author} {\bibfnamefont {P.}~\bibnamefont
  {Grassberger}}\ and\ \bibinfo {author} {\bibfnamefont {A.}~\bibnamefont
  {de~la Torre}},\ }\href@noop {} {\bibfield  {journal} {\bibinfo  {journal}
  {Ann. Phys.}\ }\textbf {\bibinfo {volume} {122}},\ \bibinfo {pages} {373}
  (\bibinfo {year} {1979})}\BibitemShut {NoStop}%
\bibitem [{\citenamefont {Cardy}\ and\ \citenamefont {Sugar}(1980)}]{cardy80}%
  \BibitemOpen
  \bibfield  {author} {\bibinfo {author} {\bibfnamefont {J.~L.}\ \bibnamefont
  {Cardy}}\ and\ \bibinfo {author} {\bibfnamefont {R.~L.}\ \bibnamefont
  {Sugar}},\ }\href@noop {} {\bibfield  {journal} {\bibinfo  {journal} {J.
  Phys. A.}\ }\textbf {\bibinfo {volume} {13}},\ \bibinfo {pages} {423}
  (\bibinfo {year} {1980})}\BibitemShut {NoStop}%
\bibitem [{\citenamefont {Bass}(1969)}]{bass69}%
  \BibitemOpen
  \bibfield  {author} {\bibinfo {author} {\bibfnamefont {F.~M.}\ \bibnamefont
  {Bass}},\ }\href@noop {} {\bibfield  {journal} {\bibinfo  {journal} {Manag.
  Sci.}\ }\textbf {\bibinfo {volume} {15}} (\bibinfo {year}
  {1969})}\BibitemShut {NoStop}%
\bibitem [{\citenamefont {Easley}\ and\ \citenamefont
  {Kleinberg}(2010)}]{easley2010}%
  \BibitemOpen
  \bibfield  {author} {\bibinfo {author} {\bibfnamefont {D.}~\bibnamefont
  {Easley}}\ and\ \bibinfo {author} {\bibfnamefont {J.}~\bibnamefont
  {Kleinberg}},\ }\href@noop {} {\emph {\bibinfo {title} {{Networks, crowds,
  and markets: Reasoning about a highly connected world}}}}\ (\bibinfo
  {publisher} {Cambridge University Press},\ \bibinfo {year}
  {2010})\BibitemShut {NoStop}%
\bibitem [{\citenamefont {Hoppensteadt}\ and\ \citenamefont
  {Izhikevich}(2012)}]{hoppenstedt12}%
  \BibitemOpen
  \bibfield  {author} {\bibinfo {author} {\bibfnamefont {F.~C.}\ \bibnamefont
  {Hoppensteadt}}\ and\ \bibinfo {author} {\bibfnamefont {E.~M.}\ \bibnamefont
  {Izhikevich}},\ }\href@noop {} {\emph {\bibinfo {title} {{Weakly connected
  neural networks}}}},\ Vol.\ \bibinfo {volume} {126}\ (\bibinfo  {publisher}
  {Springer Science \& Business Media},\ \bibinfo {year} {2012})\BibitemShut
  {NoStop}%
\bibitem [{Note1()}]{Note1}%
  \BibitemOpen
  \bibinfo {note} {The network is based on the public-use dataset from Add
  Health, a program project designed by J. Richard Udry, Peter S. Bearman, and
  Kathleen Mullan Harris, and funded by a grant from the National Institute of
  Child Health and Human Development (P01-HD31921). For data files from Add
  Health contact Add Health, Carolina Population Center, 123 W. Franklin
  Street, Chapel Hill, NC 27516-2524,
  http://www.cpc.unc.edu/addhealth.}\BibitemShut {Stop}%
\end{thebibliography}%
\bibliographystyle{apsrev4-1}
\end{document}